\documentclass{article}

\usepackage{PRIMEarxiv}

\usepackage[utf8]{inputenc} % allow utf-8 input
\usepackage[T1]{fontenc}    % use 8-bit T1 fonts
\usepackage{hyperref}       % hyperlinks
\usepackage{url}            % simple URL typesetting
\usepackage{booktabs}       % professional-quality tables
\usepackage{amsmath,amssymb,amsfonts}     % blackboard math symbols
\usepackage{eucal}
\usepackage{bm}
\usepackage{enumitem}
\usepackage{epstopdf}
\usepackage{nicefrac}       % compact symbols for 1/2, etc.
\usepackage{microtype}      % microtypography
\usepackage{lipsum}
\usepackage{fancyhdr}       % header
\usepackage{graphicx}       % graphics
\graphicspath{{media/}}     % organize your images and other figures under media/ folder

\newtheorem{theorem}{Theorem}[section]

\newtheorem{lemma}[theorem]{Lemma}
\newtheorem{remark}{Remark}
\newtheorem{definition}{Definition}

\newenvironment{proof*}{{\indent \indent \it Proof:\quad}}{\hfill $\blacksquare$\par}

\title{Modelling and Kron reduction of power flow networks in directed graphs
}

\author{
  Ruohan Wang\\
  Department of Electrical Engineering \\
  Technische Universiteit Eindhoven \\
  Eindhoven, Netherlands\\
 \texttt{r.wang@student.tue.nl} \\
   \And
  Zhiyong Sun \\
  Department of Electrical Engineering \\
  Technische Universiteit Eindhoven \\
  Eindhoven, Netherlands\\
  \texttt{z.sun@tue.nl} \\
}

\begin{document}

\maketitle
\begin{abstract}
Electrical grids are large-sized complex systems that require strong computing power for monitoring and analysis. Kron reduction is a general reduction method in graph theory and is often used for electrical circuit simplification. In this paper, we propose a novel formulation of the weighted Laplacian matrix for directed graphs. The proposed matrix is proved to be strictly equivalent to the conventionally formulated Laplacian matrix and is verified to well model a lossless DC power flow network in directed graphs. We as well present significant properties of the proposed weighted Laplacian and conditions of Kron reduction in directed graphs and in lossless DC power flow networks. The reduction method is verified via simulation models of IEEE-3, IEEE-5, IEEE-9, IEEE-14, and IEEE RTS-96 test system.
\end{abstract}

% keywords can be removed
\keywords{ Directed graphs \and Laplacian matrix \and Incidence matrix \and Kron reduction \and Schur complement \and DC power flow }

\section{\large Introduction} \label{section-Introduction}
\subsection{Background and motivations}
Large-scale systems such as electrical grids require a heavy computing workload due to their sizes and complexity. It is only natural to think of applying model reduction techniques to ease the workload. Kron reduction is a ubiquitous reduction method in electrical circuit analysis. Kron reduction is widely used in control theory and engineering to simplify and analyze large-scale systems, particularly in the design of control systems for electric power grids, aircraft, and other complex systems. It is also used in other fields, such as biology and economics, where it can be used to reduce the complexity of models and make them more tractable for analysis and simulation. Originally proposed in \cite{kron1939tensor} as purely algebraic Gaussian elimination of certain vertices in electrical circuits, Kron reduction can also be viewed from the standpoint of graph theory. By the nature of electrical circuit modeling, most of the existing model reduction work in the field of control theory is based on undirected graphs. However, in many applications including networked control systems, directed graphs arise just as naturally as undirected ones, hence before the reduction process, it is of interest to think about using directed graphs for electrical power network modeling.

\subsection{Literature review}
In this subsection, we review some existing research work on the analysis and model reduction of electrical networks. 
\par An algorithm for maximizing power flows within a power network to prevent catastrophic power outages was proposed and verified in \cite{armbruster2005power}. A parallel distributed memory structure exploiting framework which accelerates the solution of the Security Constrained Optimal Power Flow (SCOPF) problems was proposed in \cite{kardovs2019two}. Basic graph theories were used to lay the foundation for further discussion in \cite{armbruster2005power} and \cite{kardovs2019two}. However, in both papers, the main focus was the development of the proposed algorithms for a full-sized network, where the model reduction technique was not taken into consideration.
\par A novel notion termed \emph{cutset angle} was eloquently proposed by Dobson in \cite{dobson2011voltages} with the purpose of monitoring power flow network stress. The formulation of \emph{cutset angle} by Dobson could be viewed as a two-stage treatment: add a synthetic vertex being the algebraically weighted sum of all other vertices to the network, and apply Kron reduction to the network eliminating all vertices except for the synthetic vertex. Undirected graphs were used by Dobson for modelling electrical circuits in \cite{dobson2011voltages}. Similarly, the terminal voltage/current behavior of a purely linear resistive circuit was derived in \cite{willems2009behavior} by J. C. Willems and E.I. Verriest, followed by \cite{van2010characterization}, where A. van der Schaft characterized the input-output behaviors of a linear resistive circuit before and after the removal of certain vertices. The heavy usage of the symmetric weighted Laplacian of a graph was the highlight of \cite{van2010characterization}. 
\par Meanwhile, Dörfler et al. provided a detailed graph-theoretic analysis of the Kron reduction process in \cite{dorfler2010spectral}, which was followed by the application of Kron reduction on resistive circuits in \cite{dorfler2012kron}. Purely algebraic conditions that relate synchronization and transient stability of a power network were derived by Dörfler et al. in \cite{dorfler2012synchronization}. Then in \cite{dorfler2014synchronization}, Dörfler et al. further proposed analytical approaches to phase and frequency synchronization in Kron-reduced networks. In \cite{dorfler2018electrical}, Dörfler et al. surveyed both historic and recent results on electrical network analysis based on algebraic graph theory. Dörfler et al. concluded \cite{dorfler2018electrical} by a series of open questions at the intersection of algebraic graph theory and electrical networks. Based on Dörfler's work, the Kron-reduced model was used to analyze both the transient and steady-state behavior of unreduced electrical networks in \cite{caliskan2014towards}. Also based on Dörfler's work, a time-domain generalization of Kron reduction for purely resistive and inductive
networks was put forth in \cite{singh2022time}. However, despite the fact that the mentioned series of work on Kron reduction and its application on electrical networks were comprehensive and enlighting, all of them were still solely targeting undirected graphs.
\par Young et al. introduced a pairwise property of vertices that only depends on connections between the vertices in \cite{young2015new}, which is a novel generalized notion of effective resistances that apply to both undirected and directed graphs. The focus of the very paper was the development of the foundation of effective resistances for their application involving directed graphs. In \cite{sugiyama2022kron}, Sugiyama et al. extensively elaborated on Kron reduction to directed graphs. Despite that modelling electrical networks as directed graphs was briefly mentioned in \cite{young2015new} and \cite{sugiyama2022kron}, little physical interpretation of electrical networks had been covered, unfortunately.

\subsection{Contributions}
\par Contributions of this paper are summarized as follows:
\begin{itemize}
\item Modelling power flow networks in directed graphs is justified. A novel expression for the weighted directed Laplacian matrix using the graph's incidence matrix is proposed and proved to be strictly equivalent to the conventional weighted Laplacian.
\item A number of properties of the proposed weighted Laplacian matrix are analyzed, including its eigenvalue,  entry values and the existence of Schur complements. These properties are significant to model and characterize power flow networks in directed graphs. 

\item Input and output behaviors of a lossless power flow network are characterized by the proposed weighted Laplacian. I/O behaviors of the reduced network are characterized by the reduced Laplacian matrix.
\item Implementations of Kron reduction to IEEE-3, IEEE-5, IEEE-9, and IEEE-14 are successfully delivered. Numerical results of network reduction on IEEE-14 test feeder and IEEE RTS-96 are presented, showing that the proposed approach can be applied to power networks with considerable sizes.
\end{itemize}

\subsection{Organization}
\par Section \ref{section-problemformulation} gives a summary of the problem formulation of this paper. Section \ref{section-preliminatyandmath} recalls some preliminaries in matrix analysis and algebraic graph theory. Section \ref{section-mainresults} presents the formulation of the weighted Laplacian in the context of a DC power flow network and graph-theoretic analysis of Schur complements. Section \ref{section-main-results-kron} presents the graph-theoretic analysis of the Kron reduction process on DC power flow networks. Section \ref{section-numericalresults} presents numerical results of the proposed Kron reduction to an IEEE-14 test feeder and the modified IEEE RTS-96 test system. Finally, Section \ref{section-conclusionsandmore} concludes the paper and suggests future research directions.

\section{\large Problem formulation} \label{section-problemformulation}
We seek answers to these particular questions.
\begin{itemize}
\item How to model a lossless DC power flow network using a directed weighted graph?
\item What are the properties of the proposed weighted Laplacian matrix?
\item How is the proposed weighted Laplacian matrix related to the conventionally defined Laplacian matrix?
\item Does Kron reduction always exist for a directed graph?
\item Can Kron reduction always be performed to a lossless power flow network?
\item How are input-output behaviors of the original network and the reduced one related?
\end{itemize}
\par These are the major problems that motivate the work. Some were formulated during the literature review phase, and others arose inevitably during the model reduction process, which in return complemented problem formulation.

\section{\large Preliminaries} \label{section-preliminatyandmath}
\subsection{Schur complement}
Schur complement will be introduced in this section since it is the core of Kron reduction. Consider a partitioned matrix $\mathcal{M}=\left(\begin{array}{ll}
\mathcal{P} & \mathcal{Q} \\
\mathcal{R} & \mathcal{X}
\end{array}\right)$, where $\mathcal{P}, \mathcal{Q},\mathcal{R} ,\mathcal{X}$ are respectively $p\times p,p\times q,q\times p, q\times q$ sized matrices and the non-singular matrix $\mathcal{P}$ is called the leading principal sub-matrix of $\mathcal{M}$ \cite{zhang2006schur}. The term `Schur complement' of $\mathcal{P}$ was introduced by Schur: $\mathcal{M} / \mathcal{P} \triangleq \mathcal{X}-\mathcal{R} \mathcal{P}^{-1} \mathcal{Q}$. Note that Schur complement exists with respect to any non-singular sub-matrix formed with columns and rows from the original matrix. Let $\alpha,\beta$ be given index sets, which are subsets of $\left\{1,2,...,p+q\right\}$. We denote the cardinality of an index set $\alpha$ by the notation $\lvert \alpha \rvert$ and its complement by the notation $\alpha^{c}=\left\{1,2,...,p+q\right\}\setminus \alpha$. Let $\mathcal{M}\left [ \alpha,\beta \right ] $ denote the sub-matrix of $\mathcal{M}$ formed with rows indexed by $\alpha$ and columns indexed by $\beta$. If $\mathcal{M}\left [ \alpha^c,\beta^c \right ] $ is non-singular, we denote the Schur complement of $\mathcal{M}\left [ \alpha^c,\beta^c \right ]$ by $\mathcal{M}/\mathcal{M}\left [ \alpha^c,\beta^c \right ] \triangleq \mathcal{M}\left [ \alpha,\beta \right ]-\mathcal{M}\left [ \alpha,\beta^c\right ](\mathcal{M}\left [ \alpha^c,\beta^c \right ]) ^{-1} \mathcal{M}\left [ \alpha^c,\beta\right ]$. 

\subsection{Kron reduction} \label{PRE-KRON}
\begin{figure}[]
\centering
\includegraphics[width=0.8 \textwidth]{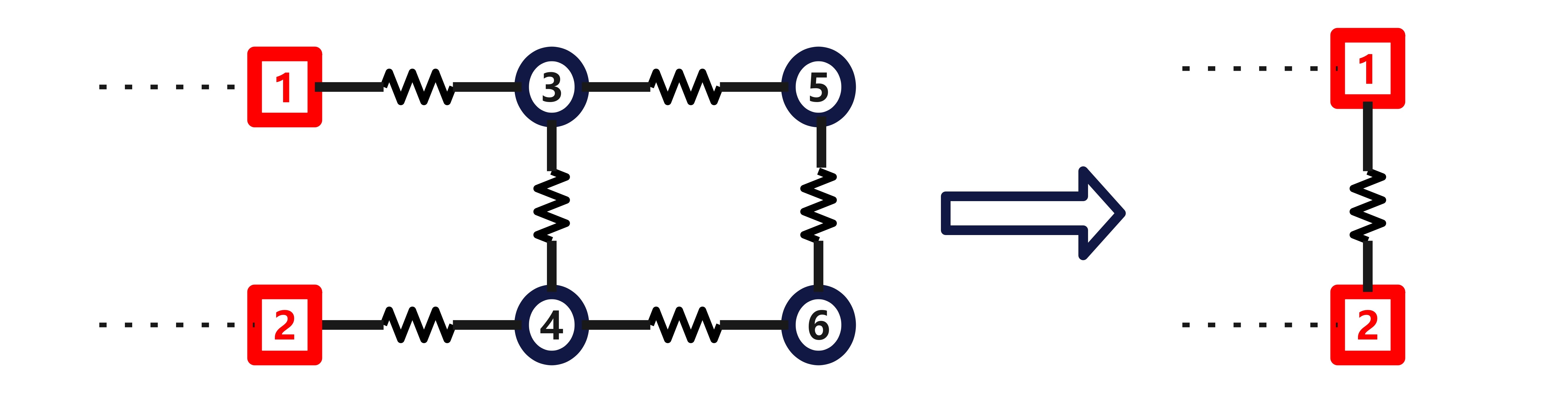}
\caption{Gaussian elimination on resistive circuits}
\label{pre-kron_circuits}
\end{figure}   
Kron reduction is a general method in graph theory for reducing the size of an electrical network by removing unimportant vertices and edges. It was first introduced by Kron as ‘Reduction Formulas’ in \cite{kron1939tensor} which was obtained through a pure Gaussian elimination procedure. 
\par Consider \textbf{a linear resistive circuit} with $n$ vertices, vertex voltages $V \in \mathbb{R}^{n \times 1}$, vertex current injections $I \in \mathbb{R}^{n \times 1}$, branch impedances $z_{ij} \ge 0$ connecting vertex $i$ and vertex $j$ and the impedance matrix $Z\in \mathbb{R}^{n \times n} $, which is a Laplacian matrix. By partitioning vertices following Kirchhoff's laws into two subsets: border vertices $\alpha \subset \left\{ 1,...,n\right\}, |\alpha| \ge2$ and inner vertices $\alpha^c = \left\{ 1,...,n\right\}\setminus \alpha$, the current-balance equations for the network can be partitioned as
\begin{align}
\left[\begin{array}{c}
V_{\alpha} \\
V_{\alpha^{c}}
\end{array}\right]=\left[\begin{array}{cc}
Z_{\alpha \alpha} & Z_{\alpha \alpha^{c}} \\
Z_{\alpha^{c} \alpha} & Z_{\alpha^{c} \alpha^{c}}
\end{array}\right]\left[\begin{array}{c}
I_{\alpha} \\
I_{\alpha^{c}}
\end{array}\right]. \label{01}
\end{align} 
Gaussian elimination of inner current injections $I_{\alpha^c}$ in (\ref{01}) gives an electrically-equivalent reduced network with border vertices $\alpha$ obeying the reduced current-balance equations
\begin{align}
V_{\alpha}+Z_{ac}V_{\alpha^c}=Z_{red}I_{\alpha} \label{02}
\end{align}  

\noindent where the reduced impedance matrix $Z_{red}$ is given by the Schur complement of $Z$ with respect to inner vertices, that is $Z_{red}=Z_{\alpha\alpha}-Z_{\alpha\alpha^c}Z_{\alpha^c\alpha^c}^{-1}Z_{\alpha^c\alpha}$, and the accompanying matrix $Z_{ac}=-Z_{\alpha\alpha^c}Z_{\alpha^c\alpha^c}^{-1}$ maps inner voltages to border voltages in the reduced network.

\par \textbf{\emph{Example}} A linear resistive circuit with 6 vertices is presented in Fig. \ref{pre-kron_circuits}. Vertices $3,4,5,6$ are \emph{inner vertices} that are only connected to other vertices within the network. Vertices $1,2$ are \emph{border vertices} that are connected to other vertices within the network \textbf{and} voltage/current sources outside the network. Current-balance equations for this network can be partitioned in the form of (\ref{01}). Gaussian elimination of inner vertices gives an electrically-equivalent reduced network, of which border vertices obey the reduced current-balance equations given by (\ref{02}). This example illustrates that both a linear resistive circuit and the reduced network eliminating all inner vertices can be described by the matrix-formed current-balance equations. $\hfill \square$

\par Similarly, consider \textbf{a lossless DC power flow network} with $n$ vertices, vertex active powers $P\in \mathbb{R}^{n\times1}$, vertex angles $\theta\in\mathbb{R}^{n\times1}$, branch susceptances $b_{ij}\ge0$ connecting vertex $i$ and vertex $j$ and the susceptance matrix $S\in\mathbb{R}^{n\times n}$ which is a Laplacian matrix. By partitioning vertices into two subsets: border vertices $\beta \subset \left\{ 1,...,n\right\}, |\beta| \ge2$ and inner vertices $\beta^c = \left\{ 1,...,n\right\}\setminus \beta$, the power-angle equation for the network can be partitioned as
\begin{align}
\left[\begin{array}{c}
P_{\beta} \\
P_{\beta^{c}}
\end{array}\right]=\left[\begin{array}{cc}
S_{\beta \beta} & S_{\beta \beta^{c}} \\
S_{\beta^{c} \beta} & S_{\beta^{c} \beta^{c}}
\end{array}\right]\left[\begin{array}{c}
\theta_{\beta} \\
\theta_{\beta^{c}}
\end{array}\right]. \label{001}
\end{align} 
\par Gaussian elimination of inner angles $\theta_{{\beta}^c}$ in (\ref{001}) gives an electrically-equivalent reduced network with border vertices $\beta$ obeying the reduced power-angle equation:
\begin{align}
P_{\beta}+S_{ac}P_{\beta^c}=S_{red}\theta_{\beta} \label{002}
\end{align}  

\noindent where the reduced susceptance matrix $S_{red}$ is given by the Schur complement of $S$ with respect to inner vertices, that is $S_{red}=S_{\beta\beta}-S_{\beta\beta^c}S_{\beta^c\beta^c}^{-1}S_{\beta^c\beta}$, and the accompanying matrix $S_{ac}=-S_{\beta\beta^c}S_{\beta^c\beta^c}^{-1}$ maps inner active powers to border active powers in the reduced network. Kron reduction will be performed mainly to \textbf{lossless DC power networks} in this paper.
\par
\begin{figure}[]
\centering
\includegraphics[width=0.8 \textwidth]{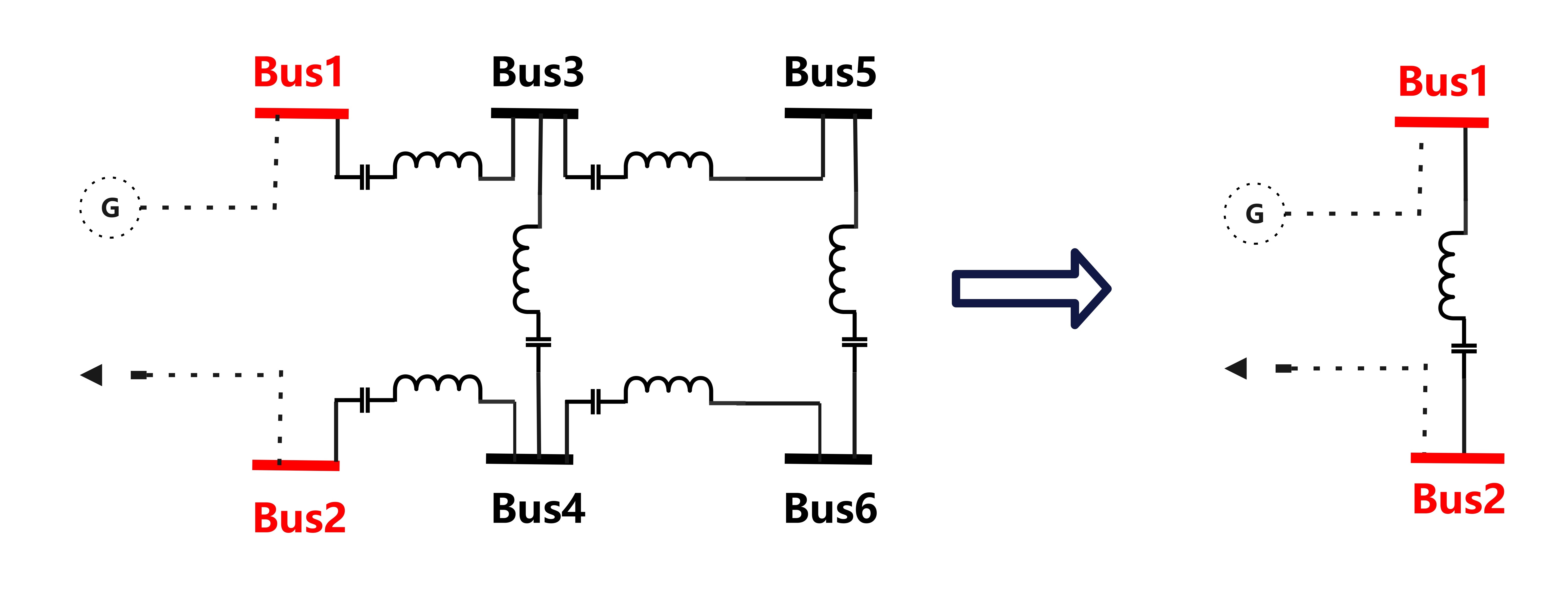}
\caption{Gaussian elimination on lossless DC power flow networks}
\label{pre-kron_pf}
\end{figure}

\textbf{\emph{Example}} A lossless DC power flow network with 6 buses/vertices is presented in Fig. \ref{pre-kron_pf}. Vertices $3,4,5,6$ are \emph{inner vertices} that are only connected to other vertices within the network. Vertices $1,2$ are \emph{border vertices} that are connected to other vertices within the network and generators/loadings outside the network. Power-angle equation for this network can be partitioned in the form of (\ref{001}). Gaussian elimination of inner vertices gives an electrically-equivalent reduced network, of which border vertices obey the reduced power-balance equations given by (\ref{002}). Similarly, this example illustrates that both a lossless DC power flow network and the reduced network eliminating all inner vertices can be described by the matrix-formed power-angle equation and that the reduction process essentially performs the Schur complement.  $\hfill \square$

 Before continuing to the next subsection, we would like to distinguish between \emph{block-by-block Kron reduction} and \emph{iterative Kron reduction}. First, we recall the definition of \emph{iterative Kron reduction}.
\begin{definition}[{Iterative Kron reduction, T. Sugiyama and K. Sato \cite{sugiyama2022kron}}]
Iterative Kron reduction associated to a weighted Laplacian matrix $\mathcal{L} \in \mathbb{R}^{n \times n}$ and indices $\left\{1,...,|\alpha| \right\}$, is a sequence of matrices $\mathcal{L}^l \in \mathbb{R}^{(n-l) \times (n-l)}$, $l \in \left\{1,...,n-|\alpha| \right\}$, which is defined as 
\begin{align}
\mathcal{L}^l=\mathcal{L}^{l-1}/\mathcal{L}^{l-1}[\left\{k_l\right\},\left\{k_l\right\}]
\end{align}
\noindent where $\mathcal{L}^0=\mathcal{L}$ and $k_l=n+1-l$.
\end{definition}

\begin{remark} \label{bbb-iterative}
\emph{Block-by-block Kron reduction} eliminates more than one vertex during each reduction process. We adopt \emph{block-by-block Kron reduction} as the main reduction method in this paper, which essentially performs the Schur complement with block sub-matrices. In contrast, \emph{iterative Kron reduction} eliminates one vertex during each reduction process. The reduction result of \emph{block-by-block Kron reduction} has been proved to be strictly equivalent to that of \emph{iterative Kron reduction} when the same vertex subset is eliminated in \cite{sugiyama2022kron}.
\end{remark}

\subsection{Directed graph, the incidence matrix and its variation} \label{Directed graph section}Consider a directed and \textbf{unweighted} graph $\mathcal{G}_d=(\mathcal{V},\varepsilon_d,\mathcal{H})$, where $\mathcal{V}$ denotes the finite vertex set, $\varepsilon_d$ denotes the directed edge set and $\mathcal{H}\in\mathbb{R}^{|\mathcal{V}|,|\mathcal{\varepsilon}_d|}$ is the corresponding unique incidence matrix. $|\mathcal{V}|$ is the number of vertices, and $|\mathcal{\varepsilon}_d|$ is the number of edges. The $(i,j) $th element $[\mathcal{H}]_{ij}$ of the incidence matrix $\mathcal{H}$ is equal to 1 if vertex $i$ is the head of the edge $j$, is equal to -1 if the vertex is the tail of edge, and 0 otherwise. The head/tail specification in the context of $a$ $DC$ $power$ $flow$ $network$ $graph$ is determined by the positioning of diode-like-functional reactive components on transmission lines, which will be elaborated in Section \ref{subsection_weighted Laplacian matrix}. Thus the incidence matrix functions as a mapping from $\varepsilon_d$ to the set of ordered pairs of $(v,w)\in \mathcal{V}^2$, with no self-loops allowed in the graph under consideration. For a given graph $\mathcal{G}_d$, we identify a subset $\mathcal{V}_{b} \subset \mathcal{V}$ as \emph{boundary vertices}. Vertices that are tails of all edges connected to them are \emph{sink} vertices. Vertices that are heads of all edges connected to them are \emph{source} vertices. \emph{Sink} and \emph{source} vertices are \emph{boundary vertices}. \emph{Boundary vertices} \textbf{cannot} be eliminated. The subset $\mathcal{V}_i=\mathcal{V}\setminus\mathcal{V}_{b}$ contains all the other vertices of the graph, being called \emph{interior vertices}. \emph{Interior vertices} \textbf{can} be eliminated. A power flow network usually includes both \emph{sink} and \emph{source}. By identifying them as \emph{boundary vertices}, we ensure the integrity of the reduced graphs.

\par Consider a directed and \textbf{weighted} graph $\mathcal{G}_d=(\mathcal{V},\varepsilon_d,\mathcal{A})$. Entries $[\mathcal{A}]_{ij}$ of the adjacency matrix $\mathcal{A}$ can be expressed in:
\begin{align}
 [\mathcal{A}]_{ij}\triangleq\left\{\begin{array}{c}
b_{k}, \text{ if }(v_i,v_j)=e_{ij}\in \varepsilon_d, b_{k} \text{ is the weight on edge } e_{ij}\\
0, \text{otherwise.}
\end{array}\right 
.
\end{align}
A diagonal degree matrix $\mathcal{D}$ corresponding to the directed and weighted graph $\mathcal{G}_d$ can be derived from the introduced adjacency matrix $\mathcal{A}$. Diagonal entries $[\mathcal{D}]_{ii}$ of the diagonal degree matrix $\mathcal{D}$ are defined as: $[\mathcal{D}]_{ii} \triangleq \sum_{j=1}^{n} [\mathcal{A}]_{ij}$.
\par Before continuing to the formulation of the weighted Laplacian matrix for a lossless power flow network, definitions of different classes of directed graphs and \emph{walk products} are given below for future reference.

\begin{definition}[{Strongly-connected graph}]
A directed graph is said to be \emph{strongly-connected} if every vertex can be reached from every other vertex.
\end{definition}

\begin{definition}[{Quasi-strongly-connected graph}]
A directed graph is said to be \emph{quasi-strongly-connected} if there exists one vertex that can reach all the other vertices in the graph. The very vertex is called \emph{root vertex}.
\end{definition}

\begin{definition}[{Walk products $(\mathcal{A}^k)_{0k}$, \cite{stuart2006digraphs}}] 
Let $\mathcal{A}$ be the $n \times n$ adjacency matrix for a given weighted directed graph $\mathcal{G}_d$. Let $(\mathcal{A}^k)_{0k}$ given by $(v_0,v_1)$, $(v_1,v_2)$, ..., $(v_{k-1},v_k)$ be a walk in $\mathcal{G}_d$. The walk product for the walk $(\mathcal{A}^k)_{0k}$ is 
\begin{align}
\prod_{j=1}^{k} [\mathcal{A}]_{j-1, j}. \label{walkproduct}
\end{align}
\end{definition}

\begin{remark}
The product given by the expression in (\ref{walkproduct}) is a generic quadrature of the $(v_0,v_k)$-entry of $\mathcal{A}^k$. The walk product $(\mathcal{A}^k)_{0k}$ is non-zero only when all quadrated elements $ [\mathcal{A}]_{j-1, j}$, $j=1,2,...,k$ are non-zero. A non-zero walk product $(\mathcal{A}^k)_{0k}$ indicates that there exists a directed path in $\mathcal{G}_d$ from $v_0$ to $v_k$.
\end{remark}

\par Next, we continue to formulate the modelling of a lossless power flow network via weighted Laplacian. Consider a graph $\mathcal{G}_d$. In the context of a DC power flow network, $\theta$ is the vector of angles \emph{at} vertices/buses which can be expressed as $\theta=\left[\theta_{1}, \theta_{2}, \ldots, \theta_{n}\right]$ where $\theta_{i}$ is the angle at the vertex $v_i$. The notation $\varphi$ denotes the vector of angle difference \emph{across} edges (between the head and the tail of the edge) of which entries $\varphi_k$ can be expressed as:

\begin{align}
\varphi_{k}=\theta_{i}-\theta_{j}
\end{align}

where $v_i$ is the head of $\text{edge}_k$ and $v_j$ is the tail of $\text{edge}_k$. $P_{edge}$ is the vector of active power flowing \emph{through} edges of which entries $P_{{edge}_k}$ can be expressed as:

\begin{align}
P_{{e d g e}_{k}}=b_{k} \varphi_{k}
\end{align}

where $b_k$ is the weight of $\text{edge}_k$. $P_v$ is the vector of active power \textbf{extractions} \emph{at} vertices/buses of which entries $P_{v_i}$ can be expressed as:

\begin{align}
P_{v_i}=\sum_{k=1}^{l} P_{\text {edge }_{k}} 
\end{align}

where the vertex $v_i$ is the head of $\text{edge}_k$ and the number of edges out of $v_i$ is $l$.

\par Define a matrix $\mathcal{H}_o$ being the variation of the incidence matrix $\mathcal{H}$ by replacing all $-1$ entries with $0$. In order to have a symmetric notation, define another matrix $\mathcal{H}_i$ being the variation of the incidence matrix $\mathcal{H}$ by replacing all $1$ entries with $0$. Since $\mathcal{H}=\mathcal{H}_o+\mathcal{H}_i$, $\mathcal{H}$ maps $P_{edge}$ to active power summations considering both \textbf{extractions and injections} \emph{at} vertices. $\mathcal{H}_i$ maps $P_{edge}$ to active power \textbf{injections} alone \emph{at} vertices. $\mathcal{H}_o$ maps $P_{edge}$ to active power \textbf{extractions} $P_v$ alone \emph{at} vertices, which will be the focus of this paper. Kirchhoff's treatment of circuit graphs is external currents entering/leaving certain vertices of the graph. The motivation of the treatment in this paper is analog to Kirchhoff's treatment of circuit graphs, which is external active power injecting/extracting to/from certain vertices of the graph. By considering both power injection and extraction, this hybrid treatment is indispensable in conventional power flow analysis. Still, it requires the articulation of $\mathcal{H}$, being the composition of $\mathcal{H}_o$ and $\mathcal{H}_i$. This conventional treatment will inevitably result in operations on undirected graphs, which is against our intentions: reduction to directed graphs. Hence in this paper we intentionally distinguish between $\mathcal{H}_o$ and $\mathcal{H}_i$, and we emphasize $\mathcal{H}_o$. According to the author's literature review, this treatment has neither been studied nor proposed.
\par Hereby we introduce our special treatment on the formulation of \emph{vertex power balance law} and \emph{angle difference law} using the incidence matrix $\mathcal{H}$ and its variation $\mathcal{H}_o$.

\par \emph{Vertex power balance law} can be given as:
\begin{align}
\mathcal{H}_{o}P_{edge}=-P_{v}. \label{vertex-power_balance_law}
\end{align}

Correspondingly, \emph{angle difference law} can be written as:
\begin{align}
\varphi=\mathcal{H}^T\theta. \label{angle-difference_law}
\end{align}
\par The formula (\ref{vertex-power_balance_law}) describes that \emph{active power extractions} at one vertex is the summation of all active powers on the edges that have their heads at the very vertex. The formula (\ref{angle-difference_law}) illustrates that the \emph{angle difference} between any two vertices can be derived from the product of the transpose of the incidence matrix and the vector of bus angle $\theta$.   

\section{\large Modelling of power flow networks and weighted Laplacian properties} \label{section-mainresults}
\subsection{Weighted Laplacian matrix} \label{subsection_weighted Laplacian matrix}
In this subsection, for a given lossless power flow network, we present the formulation of the corresponding weighting matrix $B$, the formulation of the corresponding incidence matrix $\mathcal{H}$, and subsequently, the formulation of the corresponding weighted Laplacian matrix $\mathcal{L}$. In order to streamline the formulation of problems in the context of directed graphs, we assume that the reactances of all reactive components in the network are strictly negative. Whenever there is a line with its reactance with a non-negative value, then we remove the directed edge corresponding to this reactance. Thus we may as well define the susceptance $b_{i}$ of each reactive component as the negative reciprocal of its reactance $x_{i}$, that is $b_{i}=-\frac{1}{x_{i}}>0$, for every edge $e_i$ of the network graph. Positioning of diode-like-functional reactive components determines the $orientation$ of each edge (i.e. active power is only allowed to flow through `diodes' forwardly). Define the diagonal matrix $B\triangleq diag\left\{b_1,...,b_n \right\}$. By far we have defined the diagonal weighting matrix $B$ and the incidence matrix $\mathcal{H}$ for the directed graph $\mathcal{G}_d$ in the context of a DC power flow network. Furthermore, we will throughout assume that the network graph under consideration contains at least two vertices. The following example illustrates the formulations of $\mathcal{H}$, $\mathcal{H}_o$, and $B$. 
\par
\begin{figure}[]
\centering
\includegraphics[width=0.45 \textwidth]{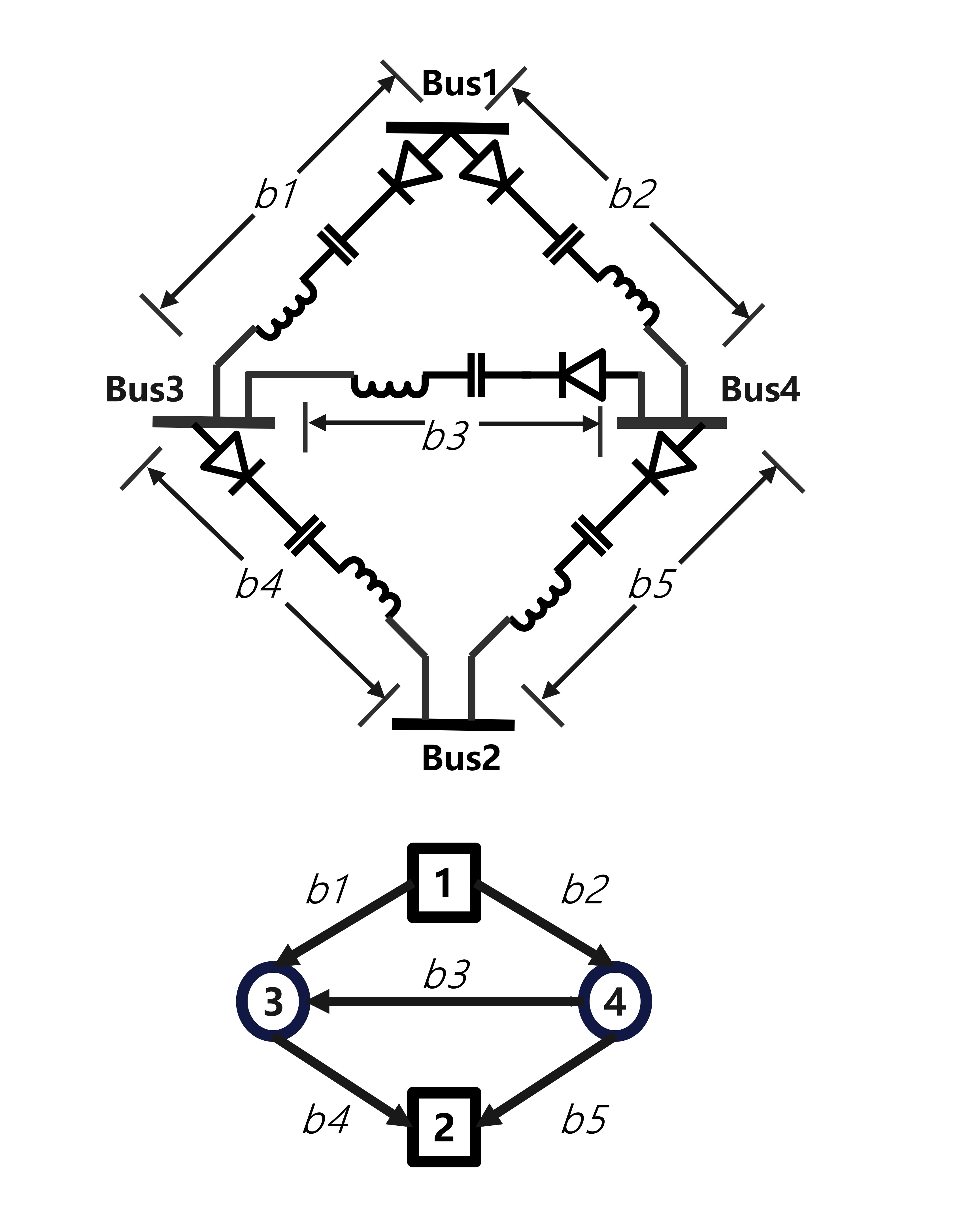}
\caption{A lossless 4-bus power flow network (upper) and its graph representation (bottom)}
\label{4_node_nored}
\end{figure}

 \textbf{\emph{Example}} Consider a lossless 4-bus power flow network; see Fig.\ref{4_node_nored} (upper). See Fig. \ref{4_node_nored} (bottom) for the corresponding graph representation of the 4-bus lossless network. Edge weights are labeled next to edges accordingly. Assume all edge susceptancs are 1. The incidence matrix $\mathcal{H}$, its variation $\mathcal{H}_o$, and the weighting diagonal matrix $B$ are:
\begin{align}
\mathcal{H}&=\left[\begin{array}{ccccc}
1 & 1 & 0 & 0 & 0 \\
0 & 0 & 0 & -1 & -1 \\
-1 & 0 & -1 & 1 & 0 \\
0 & -1 & 1 & 0 & 1 \nonumber
\end{array}\right], \\
\mathcal{H}_{o}&=\left[\begin{array}{lllll}
1 & 1 & 0 & 0 & 0 \\
0 & 0 & 0 & 0 & 0 \\
0 & 0 & 0 & 1 & 0 \\
0 & 0 & 1 & 0 & 1 \nonumber
\end{array}\right],
\\
B&=\operatorname{diag}\left\{ 1,1,1,1,1\right\} \nonumber.
\end{align}

\par This example illustrates the specification process of head/tail for every edge in the context of a DC power flow network. The specification corresponds to the formulation of the incidence matrix $\mathcal{H}$, and edge susceptances correspond to the diagonal entries of the diagonal weighting matrix $B$.  $\hfill \square$

\par To characterize the relation between vertex angles $\theta$ and vertex active power extractions $P_v$ of a lossless network, we consider a distribution of angles over its vertices, such that the corresponding angles and active power flow satisfy \emph{vertex power balance} and \emph{angle difference law}. Hence we again present the two laws given in (\ref{vertex-power_balance_law}) and (\ref{angle-difference_law}) together with the relationship between active power \emph{through} edges and the angle differences \emph{across} the edge:
\begin{align}
\varphi&=\mathcal{H}^T\theta, \label{10}
\\
P_{egde}&=-B\varphi, \label{11}
\\
-P_{v}&=\mathcal{H}_{o}P_{edge}. \label{12}
\end{align}

\par Replacing $P_{edge}$ in (\ref{12}) with its expression (\ref{11}), and replacing $\varphi$ with its expression in (\ref{10}), we have:

\begin{align}
P_v=\mathcal{H}_{o}B\mathcal{H}_{}^T\theta=\mathcal{L}\theta \label{Pv=Ltheta}
\end{align}

\noindent where $\mathcal{L}=\mathcal{H}_{o}B\mathcal{H}_{}^T$.
\par We now formally introduce our definition of the weighted Laplacian matrix.
\begin{definition}[{Weighted Laplacian matrix}] \label{weightedLaplacian}
For any directed graph with the incidence matrix $\mathcal{H}$ and the weighting diagonal matrix $B$, the square matrix $\mathcal{H}_{o}B\mathcal{H}_{}^T$ is defined as the \emph{weighted Laplacian matrix} $\mathcal{L}$ of the graph.
\end{definition}

 The weighted Laplacian matrix features many important properties, which will be elaborated on in Section \ref{section-mainresults} and lay the foundation of the characterization of input-output behavior of $any$ lossless network in Section \ref{section-main-results-kron}. Every theorem and lemma introduced in the sequel will be equipped with proof and an example for readers' understanding.
 
\subsection{Weighted Laplacians of directed graphs}
\par This subsection will present several important properties of the weighted Laplacian matrix $\mathcal{L}$ of a given directed graph $\mathcal{G}_d$. First, we present a theorem for generalized directed graphs.
\begin{theorem} \label{1sttheorem}
Consider a directed graph $\mathcal{G}_d$ with incidence matrix $\mathcal{H}$ and its variation $\mathcal{H}_{o}$. Let $B$ be a positive definite diagonal weighting matrix, of which the dimension is equal to the number of edges. Then the weighted Laplacian matrix $\mathcal{L}=\mathcal{H}_{o}B\mathcal{H}_{}^T$ has the following properties
\begin{enumerate}
\item The weighted Laplacian $\mathcal{L}$ is asymmetric, having all eigenvalues with non-negative real parts, and dependent on the orientation of the graph. 
\item The weighted Laplacian $\mathcal{L}$ has non-negative diagonal entries, and non-positive off-diagonal entries. 
\item  The weighted Laplacian $\mathcal{L}$ has zero row sums. The vector $\bm{1}$ is in the right nullspace of $\mathcal{L}$.
\end{enumerate}

\label{theorem-Laplacian-matrix-properties}
\end{theorem}

\begin{proof*} For the proof of \emph{Theorem} \ref{theorem-Laplacian-matrix-properties}, we aim to prove that our definition of the weighted Laplacian matrix is strictly equivalent to the conventional definition: $\mathcal{L}_{conv}\triangleq\mathcal{D}-\mathcal{A}$. Consider a directed graph $\mathcal{G}_d$ with its corresponding incidence matrix $\mathcal{H}$ and weighting diagonal matrix $B$. Entries $[\mathcal{L}]_{ij}$ of our novel definition of the weighted Laplacian matrix $\mathcal{L}$ can be expressed in: 

\begin{align}
\begin{array}{l}
i \neq j:[\mathcal{L}]_{i j} \triangleq\left\{\begin{array}{cl}
-b_{k}, & \text { if }\left(v_{i}, v_{j}\right)=e_{i j} \in \varepsilon_{d}, \\
& b_{k} \text { is the weight on edge } e_{i j} \\
0, & \text { if }\left(v_{i}, v_{j}\right)=e_{i j} \notin \varepsilon_{d}
\end{array}\right. \\ \\
i=j:[\mathcal{L}]_{i i} \triangleq-\sum_{p=1, p \neq i}^{n}[\mathcal{L}]_{i p}.  \label{novelLaplaciandef}
\end{array} 
\end{align}

Now recall the conventional Laplacian matrix definition for a directed graph: $\mathcal{L}_{conv}\triangleq\mathcal{D}-\mathcal{A}$, where $\mathcal{A}$ is the adjacency matrix, entries of which have been declared in Section \ref{Directed graph section}, and $\mathcal{D}$ is the diagonal degree matrix. Hence, entries $[\mathcal{L}_{conv}]_{ij}$ of the conventional Laplacian matrix $\mathcal{L}_{conv}$ can be expressed as: 
\begin{align}
\begin{array}{l}
i \neq j:[\mathcal{L}_{conv}]_{i j} \triangleq\left\{\begin{array}{cl}
-b_{k}, & \text { if }\left(v_{i}, v_{j}\right)=e_{i j} \in \varepsilon_{d}, \\
& b_{k} \text { is the weight on edge } e_{i j} \\
0, & \text { if }\left(v_{i}, v_{j}\right)=e_{i j} \notin \varepsilon_{d}
\end{array}\right. \\ \\
i=j:[\mathcal{L}_{conv}]_{i i} \triangleq-\sum_{p=1, p \neq i}^{n}[\mathcal{L}_{conv}]_{i p}.  \label{convLaplaciandef}
\end{array}
\end{align}

Observing (\ref{novelLaplaciandef}) and (\ref{convLaplaciandef}) it is evident that for a directed weighted graph $\mathcal{G}_d$, our definition of the weighted Laplacian: $\mathcal{H}_o B \mathcal{H}^{T}$ is strictly equivalent to the conventional definition: $\mathcal{D}-\mathcal{A}$. Since $\mathcal{L}_{conv}$ is known to be asymmetric, having all eigenvalues with non-negative real parts, we conclude the proof of \emph{Theorem} \ref{1sttheorem}.1. 
\par Since $b_k$ is always positive as defined in Section \ref{subsection_weighted Laplacian matrix}, off-diagonal entries of $\mathcal{L}$ are always non-positive. Observing the definition for diagonal entries of $[\mathcal{L}]_{ii}$ in (\ref{novelLaplaciandef}), it is evident that $\mathcal{L}$ has non-negative diagonal entries and zero row sums. Hence the proofs of \emph{Theorem} \ref{1sttheorem}.2 and \emph{Theorem} \ref{1sttheorem}.3 are concluded.
\end{proof*}

\par \textbf{\emph{Example}} For the directed graph in Fig. \ref{4_node_nored} (bottom), we assign edge weights $\left\{b_1,b_2,b_3,b_4,b_5\right\}$ as $\left\{1,2,3,4,5\right\}$. Then the incidence matrix $\mathcal{H}$, its variation $\mathcal{H}_{o}$, the diagonal matrix $B$, its weighted Laplacian $\mathcal{L}$ and its eigenvalues are: 
\begin{footnotesize}
\begin{align}
\begin{array}{rlrl}
\mathcal{H} & =\left[\begin{array}{ccccc}
1 & 1 & 0 & 0 & 0 \\
0 & 0 & 0 & -1 & -1 \\
-1 & 0 & -1 & 1 & 0 \\
0 & -1 & 1 & 0 & 1
\end{array}\right], & \mathcal{H}_{o}=\left[\begin{array}{ccccc}
1 & 1 & 0 & 0 & 0 \\
0 & 0 & 0 & 0 & 0 \\
0 & 0 & 0 & 1 & 0 \\
0 & 0 & 1 & 0 & 1
\end{array}\right], \\ \\
B & =\left[\begin{array}{ccccc}
1 & 0 & 0 & 0 & 0 \\
0 & 2 & 0 & 0 & 0 \\
0 & 0 & 3 & 0 & 0 \\
0 & 0 & 0 & 4 & 0 \\
0 & 0 & 0 & 0 & 5
\end{array}\right], & \mathcal{L}=\left[\begin{array}{cccc}
3 & 0 & -1 & -2 \\
0 & 0 & 0 & 0 \\
0 & -4 & 4 & 0 \\
0 & -5 & -3 & 8
\end{array}\right], 
\end{array}\nonumber
\end{align}
~
\begin{align}
\operatorname{eig}(\mathcal{L})&=\{3,8,4,0\} .
\nonumber
\end{align}
\end{footnotesize}
\par Correspondingly the adjacency matrix $\mathcal{A}$, the degree matrix $\mathcal{D}$, and its conventional Laplacian matrix $\mathcal{L}_{conv}$ are:
\begin{small}
\begin{align}
\mathcal{A}&=\left[\begin{array}{llll}
0 & 0 & 1 & 2 \\
0 & 0 & 0 & 0 \\
0 & 4 & 0 & 0 \\
0 & 5 & 3 & 0
\end{array}\right], \quad \mathcal{D}=\left[\begin{array}{llll}
3 & 0 & 0 & 0 \\
0 & 0 & 0 & 0 \\
0 & 0 & 4 & 0 \\
0 & 0 & 0 & 8
\end{array}\right],\nonumber \quad \\
\mathcal{L}_{\text {conv }}&=\left[\begin{array}{cccc}
3 & 0 & -1 & -2 \\
0 & 0 & 0 & 0 \\
0 & -4 & 4 & 0 \\
0 & -5 & -3 & 8
\end{array}\right]. \nonumber
\end{align}
\end{small}
In this example, it holds that $\mathcal{D}-\mathcal{A}=\mathcal{H}_{o}B\mathcal{H}^T$. This example presents that for a given directed graph, our definition of the Laplacian matrix is strictly equivalent to the conventional definition and that the Laplacian possesses all properties as stated in \emph{Theorem} \ref{1sttheorem}.  $\hfill \square$

\par We then introduce our lemmas on the properties of the weighted Laplacians of different classes of directed graphs.
\par Before continuing, we recall the definition for \emph{reachable subset} and \emph{Lemma} 3.2 in \cite{sugiyama2022kron} on the existence of Schur complement (with
notations changed to match this paper) for future reference:
\begin{definition}[{Reachable subset, T. Sugiyama and K. Sato \cite{sugiyama2022kron}}]
  Let $\mathcal{G}_d=(\mathcal{V},\varepsilon_d,\mathcal{H})$ be a directed and weighted graph with diagonal weighting matrix $B$ and $\mathcal{V}_\alpha \subset \mathcal{V}$ be a proper subset of vertices with $|\mathcal{V}_\alpha| \ge2$. $\mathcal{V}_{{\alpha}^c}=\mathcal{V}\setminus \mathcal{V}_{\alpha}$. We refer to $\mathcal{V}_\alpha \subset  \mathcal{V}$ as a \emph{reachable subset} of $\mathcal{G}_d$ if for \textbf{any} $v_i \in \mathcal{V}_{\alpha^c}$, there exist a vertex $v_j\in \mathcal{V}_\alpha$ and a path in $\mathcal{G}_d$ from $v_i$ to $v_j$.
\end{definition}

\begin{lemma} [{Existence of Schur complement, T. Sugiyama and K. Sato \cite{sugiyama2022kron}}] Let $\mathcal{G}_d=(\mathcal{V},\varepsilon_d,\mathcal{H})$ be a directed and weighted graph with diagonal weighting matrix $B$ and $\mathcal{V}_\alpha \subset \mathcal{V}$ be a proper subset of vertices with $|\mathcal{V}_\alpha| \ge2$. $\mathcal{V}_{{\alpha}^c}=\mathcal{V}\setminus \mathcal{V}_{\alpha}$. Then, the Schur complement of $\mathcal{L}$ with respect to the sub-matrix consisting of columns and rows corresponding to vertices $\mathcal{V}_{\alpha}$ exists \textbf{if and only if} $\mathcal{V}_\alpha$ is a reachable subset of $\mathcal{G}_d$. \label{lemma-sugiyama}
\end{lemma}

\begin{lemma} 
If the graph $\mathcal{G}_d$ is strongly-connected, then 
\begin{enumerate}
\item All diagonal entries of $\mathcal{L}$ are positive.
\item All Schur complements of $\mathcal{L}$ exist.
\end{enumerate}

\label{SC-LAPLACIAN}
\end{lemma}

\begin{proof*} 
\begin{enumerate}
\item For every vertex $v_i \in \mathcal{V}$ there exits at least one edge of which $v_i$ is the head, featuring a negative $[\mathcal{L}]_{ij}, i \ne j$, therefore all diagonal entries $[\mathcal{L}]_{ii}$ are positive. Hence the proof for \emph{Lemma} \ref{SC-LAPLACIAN}.1 is concluded.
\item Schur complements of $\mathcal{L}$ with respect to sub-matrices consisting of rows and columns corresponding to $\mathcal{V}_{\alpha}$ exist for any vertex subset $\mathcal{V}_{\alpha}$. In the case of a strong-connected graph, any vertex subset $\mathcal{V}_{\alpha}$ is always a \emph{reachable subset} to $\mathcal{V}_{{\alpha}^c}$. Hence by referring to \emph{Lemma}~\ref{lemma-sugiyama} we conclude the proof for \emph{Lemma} \ref{SC-LAPLACIAN}.2.
\end{enumerate}
\end{proof*}
\begin{figure}[]
\centering
\includegraphics[width=0.45\textwidth]{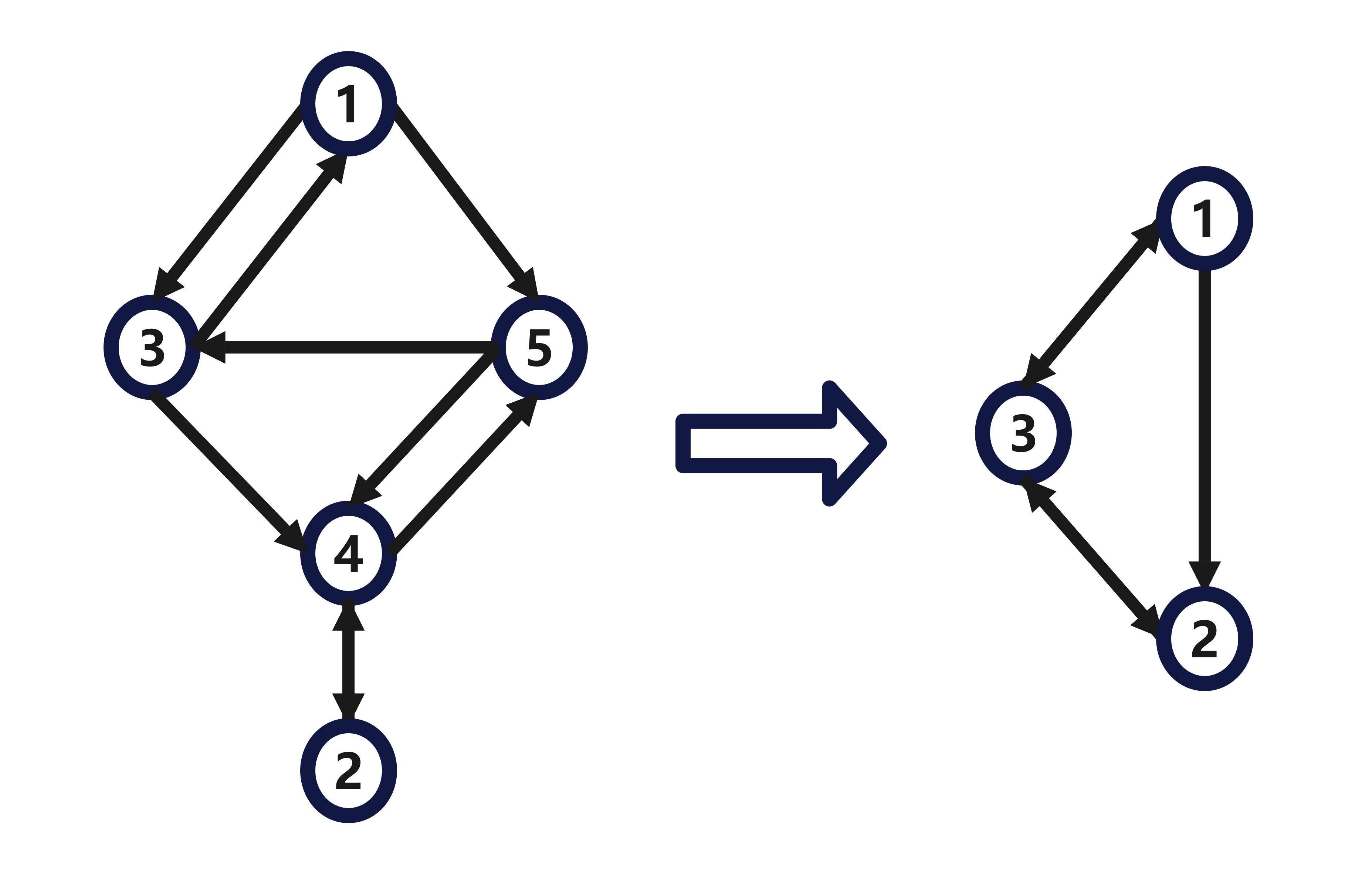}
\caption{Illustration of Kron reduction to a strongly-connected graph (Left: original graph. Right: reduced graph with vertices $4,5$ eliminated). Edge weights are omitted for simplicity.}
\label{SC_graph}
\end{figure}

\par \textbf{\emph{Example}} Consider a strongly-connected graph in Fig. \ref{SC_graph}. Assume all edge weights are 1 for simplicity. The weighted Laplacian of the original graph is:
\begin{small}
\begin{align}
\mathcal{L}=\left[\begin{array}{ccccc}
2 & 0 & -1 & 0 & -1 \\
0 & 1 & 0 & -1 & 0 \\
-1 & 0 & 2 & -1 & 0 \\
0 & -1 & 0 & 2 & -1 \\
0 & 0 & -1 & -1 & 2
\end{array}\right]. \nonumber
\end{align}
\end{small}
\par All diagonal entries of $\mathcal{L}$ are positive, and all Schur complements of $\mathcal{L}$ with respect to any sub-matrix corresponding to $\mathcal{V}_{\alpha}$ exist. In Fig. \ref{SC_graph}, vertices $4,5$ are eliminated during the reduction. The corresponding Schur complement is $\mathcal{L}_{red}$: 
\begin{small}
\begin{align}
\mathcal{L}_{\text {red }}=\left[\begin{array}{ccc}
2 & -0.333 & -1.667 \\
0 & 0.333 & -0.333 \\
-1 & -0.667 & 1.667
\end{array}\right] \nonumber .
\end{align}
\end{small}
\par The reduced weighted Laplacian $\mathcal{L}_{red}$ corresponds to the reduced graph in Fig. \ref{SC_graph} (right). This example illustrates that all diagonal entries of the weighted Laplacian matrix $\mathcal{L}$ corresponding to a strongly-connected graph $\mathcal{G}_d$ are positive. Furthermore, with respect to the sub-matrix consisting of rows and columns corresponding to any chosen vertex subset, the Schur complement of the weighted Laplacian exits.  $\hfill \square$
\par Next we introduce the properties of the weighted Laplacian of a quasi-strongly-connected graph.     
\begin{remark}
For readers' understanding, we use notations \emph{retained vertices} $\mathcal{V}_\alpha$ and \emph{eliminated vertices} $\mathcal{V}_{\alpha^c}$ which shall be formally introduced in Section \ref{subsection-vertex-class} here in \emph{Lemma} \ref{QSC-GRAPH-LAPLACIAN}. The Schur complement of $\mathcal{L}$ stated in \emph{Lemma} \ref{QSC-GRAPH-LAPLACIAN} is with respect to the sub-matrix consisting of rows and columns corresponding to \emph{retained vertices}.
\end{remark}
\begin{lemma}
If the graph $\mathcal{G}_d$ is quasi-strongly-connected, and $\mathcal{G}_d$ consists of \emph{sink} vertices, then the following statements hold.
\begin{enumerate}
\item The diagonal entries of $\mathcal{L}$ corresponding to \emph{sink} vertices are $0$, and all other diagonal entries are positive.
\item Consider the Schur complement of $\mathcal{L}$ with respect to the sub-matrix consisting of rows and columns corresponding to \emph{retained vertices} $\mathcal{V}_\alpha$. The Schur complement exists if and only if the subset of \emph{retained vertices} includes \textbf{the entire \emph{sink} vertices}.
\end{enumerate} 
 \label{QSC-GRAPH-LAPLACIAN}
\end{lemma}
\begin{proof*} 
\begin{enumerate}
\item Since \emph{sink} vertices are vertices that are tails of all edges connected to them, every other vertex not being a \emph{sink} has a positive out-degree. The diagonal entries of $\mathcal{L}$ indicate out-degrees of corresponding vertices. Hence the proof for \emph{Lemma} \ref{QSC-GRAPH-LAPLACIAN}.1 is concluded.
\item For the proof of \emph{Lemma} \ref{QSC-GRAPH-LAPLACIAN}.2, we first prove that the Schur complement exists if the subset of \emph{retained vertices} includes \textbf{the entire \emph{sink} vertices}. Since \textbf{the entire \emph{sink} vertices} are included in the subset $\mathcal{V}_{\alpha}$, there always exists a directed path in $\mathcal{G}_d$ starting at any vertex in $\mathcal{V}_{{\alpha}^c}$ and ending at a \emph{sink} vertex in $\mathcal{V}_{\alpha}$. Therefore $\mathcal{V}_{\alpha}$ is always a \emph{reachable subset} of $\mathcal{G}_d$ for $\mathcal{V}_{{\alpha}^c}$. According to \emph{Lemma} \ref{lemma-sugiyama}, the Schur complement exists if the subset of \emph{retained vertices} includes \textbf{the entire \emph{sink} vertices}. Hence we conclude the proof for the first part of \emph{Lemma} \ref{QSC-GRAPH-LAPLACIAN}.2. 
\par We then prove that the Schur complement does not exist if the subset of \emph{retained vertices} does not include \textbf{the entire \emph{sink} vertices}. Since \emph{sink} vertices are vertices that are tails of all edges connected to them, there exists no directed path in $\mathcal{G}_d$ starting at one vertex of $\mathcal{V}_{{\alpha}^c}$ and ending at any \emph{sink} vertex of $\mathcal{V}_{\alpha}$. Therefore, $\mathcal{V}_{\alpha}$ is \textbf{never} a \emph{reachable subset} of $\mathcal{G}_d$ for $\mathcal{V}_{{\alpha}^c}$. Referring to \emph{Lemma} \ref{lemma-sugiyama}, the Schur complement does not exist if the subset of \emph{retained vertices} does not include \textbf{the entire \emph{sink} vertices}. Hence the proof for \emph{Lemma} \ref{QSC-GRAPH-LAPLACIAN}.2 is concluded. Hereby we can claim that the Schur complement of $\mathcal{L}$ with respect to the sub-matrix consisting of rows and columns corresponding to \emph{retained vertices} $\mathcal{V}_\alpha$ exists if and only if the subset of \emph{retained vertices} includes \textbf{the entire \emph{sink} vertices}.
\end{enumerate}
\end{proof*}
\par
\begin{figure}[]
\centering
\includegraphics[width=0.35\textwidth]{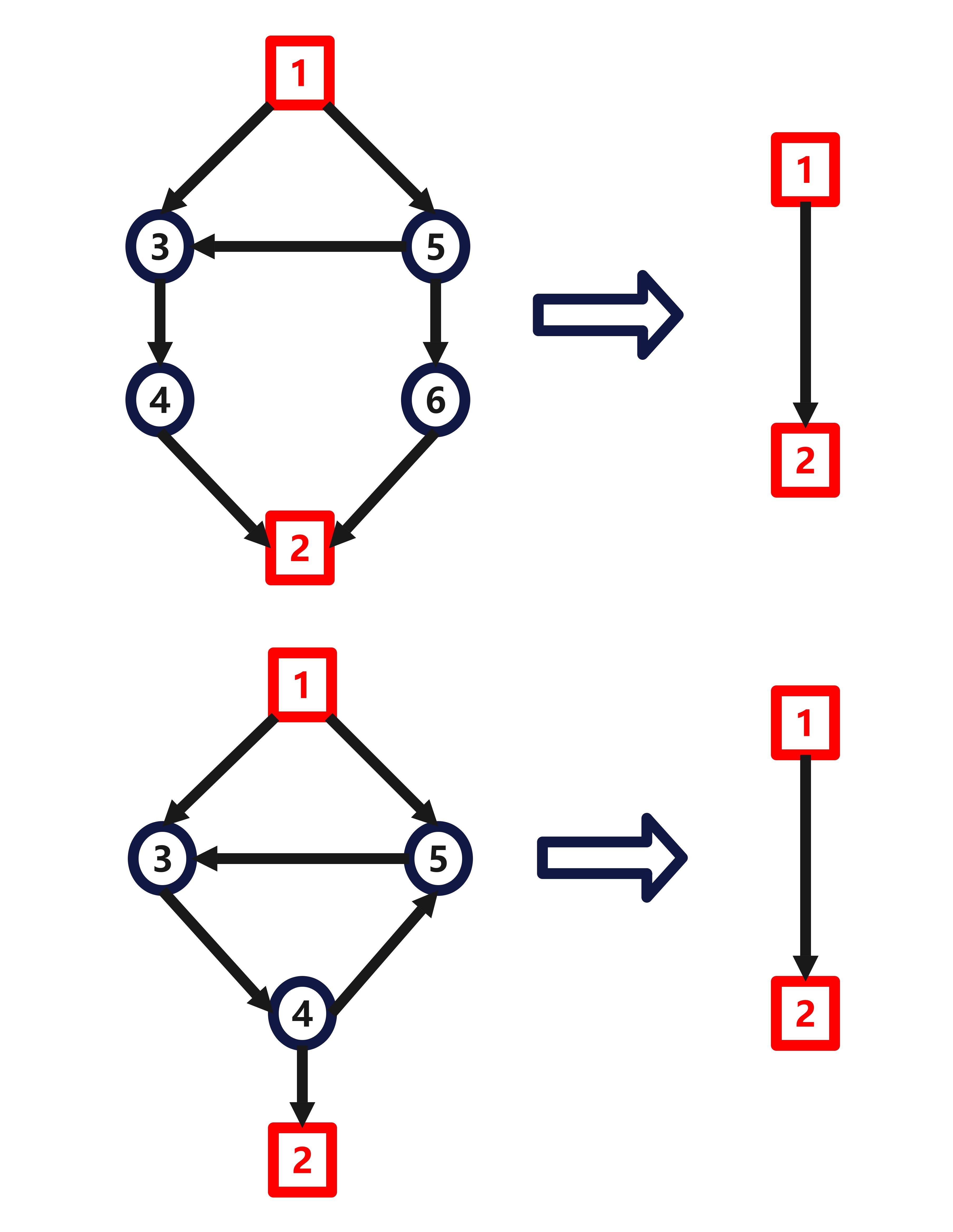}
\caption{Illustration of Kron reduction to an acyclic quasi-strongly-connected graph (upper) and a cyclic quasi-strongly-connected graph (bottom). Edge weights are omitted for simplicity. \emph{Boundary vertices} are marked in red.}
\label{sink_source_acyclic_cyclic}
\end{figure}

\textbf{\emph{Example}} For the quasi-strongly-connected graphs in Fig.~\ref{sink_source_acyclic_cyclic} (left), assuming all edge weights are 1 for simplicity, the weighted Laplacian $\mathcal{L}_{acy}$ of the acyclic graph and the weighted Laplacian $\mathcal{L}_{cyc}$ of the cyclic graph are: 
\begin{small}
\begin{align*}
\mathcal{L}_{a c y}&=\left[\begin{array}{cccccc}
2 & 0 & -1 & 0 & -1 & 0 \\
0 & 0 & 0 & 0 & 0 & 0 \\
0 & 0 & 1 & -1 & 0 & 0 \\
0 & -1 & 0 & 1 & 0 & 0 \\
0 & 0 & -1 & 0 & 2 & -1 \\
0 & -1 & 0 & 0 & 0 & 1
\end{array}\right], \quad  
\end{align*}

\begin{align}
\mathcal{L}_{c y c}&=\left[\begin{array}{ccccc}
2 & 0 & -1 & 0 & -1 \\
0 & 0 & 0 & 0 & 0 \\
0 & 0 & 1 & -1 & 0 \\
0 & -1 & 0 & 2 & -1 \\
0 & 0 & -1 & 0 & 1
\end{array}\right] . \nonumber
\end{align}
\end{small}
\par All diagonal entries of both $\mathcal{L}_{acy}$ and $\mathcal{L}_{cyc}$ except for the \emph{boundary vertex} $2$ are positive. Except for the Schur complement of sub-matrix corresponding to the \emph{boundary vertex} $2$, all the other Schur complements of $\mathcal{L}$ exist. In Fig. \ref{sink_source_acyclic_cyclic}, all \emph{interior vertices} are eliminated during the reduction ($3,4,5,6$ for the upper graph, $3,4,5$ for the bottom graph). The corresponding Schur complements are $\mathcal{L}_{acy\;red}$ and  $\mathcal{L}_{cyc\;red}$:
\begin{small}
\begin{align}
\mathcal{L}_{acy\;red}=\left[\begin{array}{cc}
2 & -2 \\
0 & 0
\end{array}\right], \quad \mathcal{L}_{cyc\;red}=\left[\begin{array}{cc}
2 & -2 \\
0 & 0
\end{array}\right]. \nonumber
\end{align}
\end{small}
\par This example illustrates that except for the diagonal entries corresponding to \emph{sink} vertices, all other diagonal entries of the weighted Laplacian $\mathcal{L}$ of a quasi-strongly-connected graph $\mathcal{G}_d$ are positive (both for cyclic and acyclic graphs). It also illustrates that the Schur complement of $\mathcal{L}$ with respect to sub-matrix consisting of rows and columns corresponding to vertices (of which the subset includes \textbf{the entire \emph{sink} vertices}) exists (both for cyclic and acyclic graphs).  $\hfill \square$

\subsection{Directed graphs corresponding to weighted Laplacian matrices}

\begin{figure}[]
\centering
\includegraphics[width=0.35\textwidth]{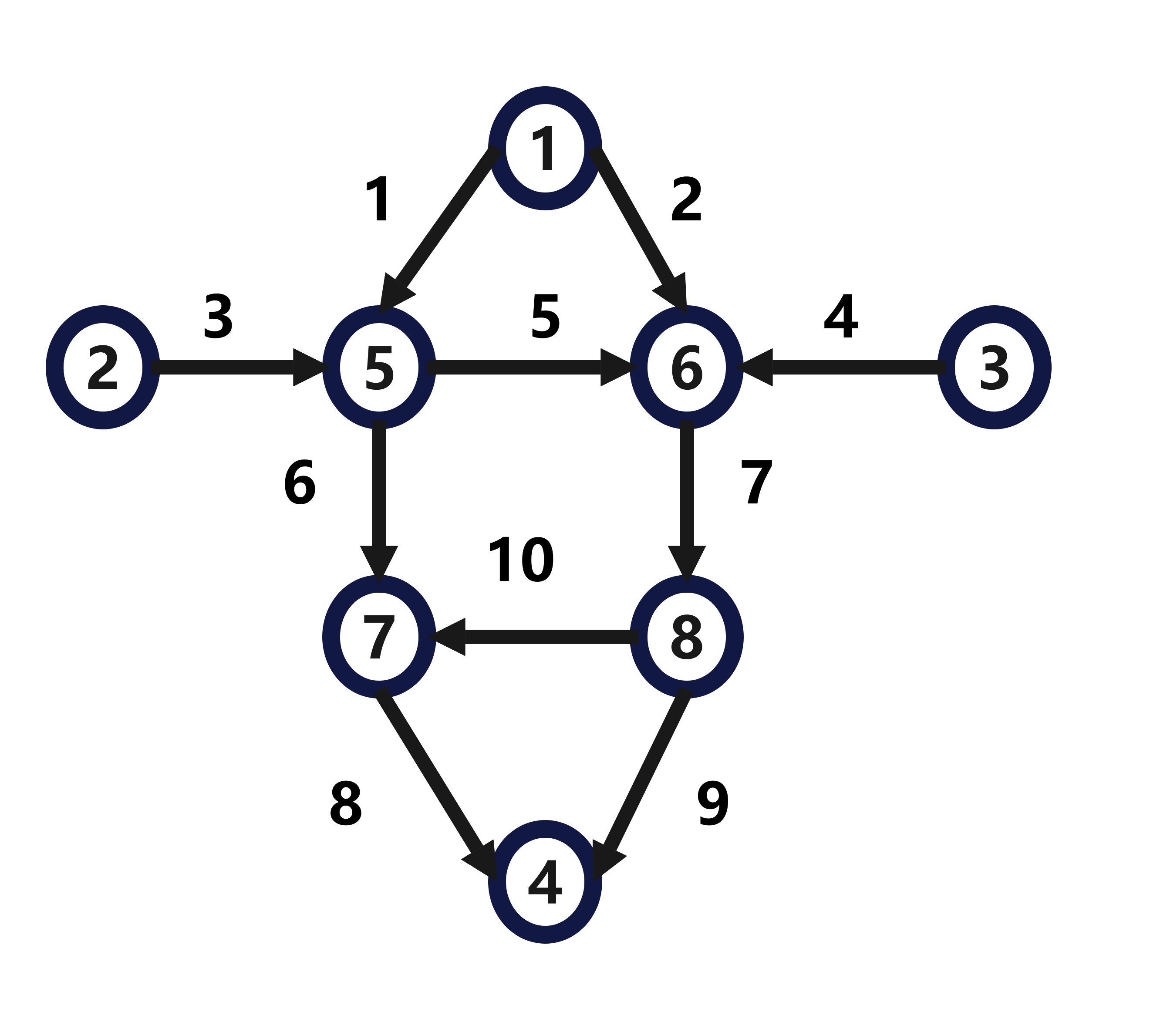}
\caption{Example of a directed graph corresponding to the Laplacian matrix given in \emph{Theorem} \ref{theorem-graph-to-Laplacian} (edge weights marked next to edges)}
\label{theorem4_6example}
\end{figure}
\par In this section, we will present that there exist directed graphs corresponding to given weighted Laplacians and reduced weighted Laplacians. We will also present that certain properties of the corresponding graph are preserved during the reduction process.
\begin{theorem} \label{theorem-graph-to-Laplacian}
Consider an asymmetric matrix $\mathcal{L}$ having all eigenvalues with non-negative real parts, non-negative diagonal entries, non-positive off-diagonal entries, and zero row sums. Then 
\begin{enumerate}
\item $\mathcal{L}$ corresponds to the Laplacian matrix of a directed weighted graph. 
\item $\mathcal{L}$ can be written as $\mathcal{L}=\mathcal{H}_{o}B\mathcal{H}_{}^T$, with $\mathcal{H}$ the incidence matrix of the corresponding graph, $\mathcal{H}_{o}$ being the appointed variation of $\mathcal{H}$, and $B$ a positive definite diagonal matrix of the corresponding graph.
\end{enumerate}

\end{theorem}
\begin{proof*}
\begin{enumerate}
\item Consider \emph{Theorem} \ref{theorem-graph-to-Laplacian} as the reverse statement of \emph{Theorem} \ref{1sttheorem}. Then for every asymmetric matrix $\mathcal{L}$ with properties stated in \emph{Theorem} \ref{theorem-graph-to-Laplacian}, there exists a weighted directed graph corresponding to it. 
\item The matrix $\mathcal{L}$ can be written as $\mathcal{L}=\mathcal{D}-\mathcal{A}$, where $\mathcal{D}$ is the graph's degree matrix and $\mathcal{A}$ is the graph's adjacency matrix. Recall that in the proof for \emph{Theorem}~\ref{1sttheorem}, we proved that $\mathcal{D}-\mathcal{A}=\mathcal{H}_{o}B\mathcal{H}_{}^T$. Hence we can declare that $\mathcal{L}$ can be written as $\mathcal{L}=\mathcal{H}_{o}B\mathcal{H}_{}^T$ with $\mathcal{H}$ the incidence matrix of the corresponding graph, $\mathcal{H}_{o}$ being the appointed variation of $\mathcal{H}$, and $B$ a positive definite diagonal matrix of the corresponding graph.
\end{enumerate}
\end{proof*}
\textbf{\emph{Example}} For the following weighted asymmetric Laplacian matrix $\mathcal{L}$, there exists a directed graph corresponding to it; see Fig. \ref{theorem4_6example}. 
\begin{small}

\begin{align*}
\mathcal{L}  =\left[\begin{array}{cccccccc}
3 & 0 & 0 & 0 & -1 & -2 & 0 & 0 \\
0 & 3 & 0 & 0 & -3 & 0 & 0 & 0 \\
0 & 0 & 4 & 0 & 0 & -4 & 0 & 0 \\
0 & 0 & 0 & 0 & 0 & 0 & 0 & 0 \\
0 & 0 & 0 & 0 & 11 & -5 & -6 & 0 \\
0 & 0 & 0 & 0 & 0 & 7 & 0 & -7 \\
0 & 0 & 0 & -8 & 0 & 0 & 8 & 0 \\
0 & 0 & 0 & -9 & 0 & 0 & -10 & 19
\end{array}\right].  
\end{align*}

\end{small}

 \par The corresponding incidence matrix $\mathcal{H}$ and the diagonal matrix $B$ are:

 \begin{small}
\begin{align}
\mathcal{H}  &=\left[\begin{array}{cccccccccc}
1 & 1 & 0 & 0 & 0 & 0 & 0 & 0 & 0 & 0 \\
0 & 0 & 0 & 1 & 0 & 0 & 0 & 0 & 0 & 0 \\
0 & 0 & 0 & 0 & 1 & 0 & 0 & 0 & 0 & 0 \\
0 & 0 & 0 & 0 & 0 & 0 & 0 & 0 & -1 & -1 \\
-1 & 0 & 1 & -1 & 0 & 1 & 0 & 0 & 0 & 0 \\
0 & -1 & -1 & 0 & -1 & 0 & 1 & 0 & 0 & 0 \\
0 & 0 & 0 & 0 & 0 & -1 & 0 & -1 & 1 & 0 \\
0 & 0 & 0 & 0 & 0 & 0 & -1 & 1 & 0 & 1 \nonumber
\end{array}\right], \\ 
B&=\operatorname{diag}\left\{ 1,2,5,3,4,6,7,10,8,9\right\} \nonumber.
\end{align}
\end{small}
\par This example illustrates that an asymmetric matrix with the properties stated in \emph{Theorem} \ref{theorem-graph-to-Laplacian} always corresponds to a weighted directed graph and can be written as $\mathcal{L}=\mathcal{H}_o B \mathcal{H}^T$ with edge directions encoded in the incidence matrix $\mathcal{H}$, edge weights encoded in the weighting diagonal matrix $B$. $\hfill \square$

\begin{figure}[]
\centering
\includegraphics[width=0.45\textwidth]{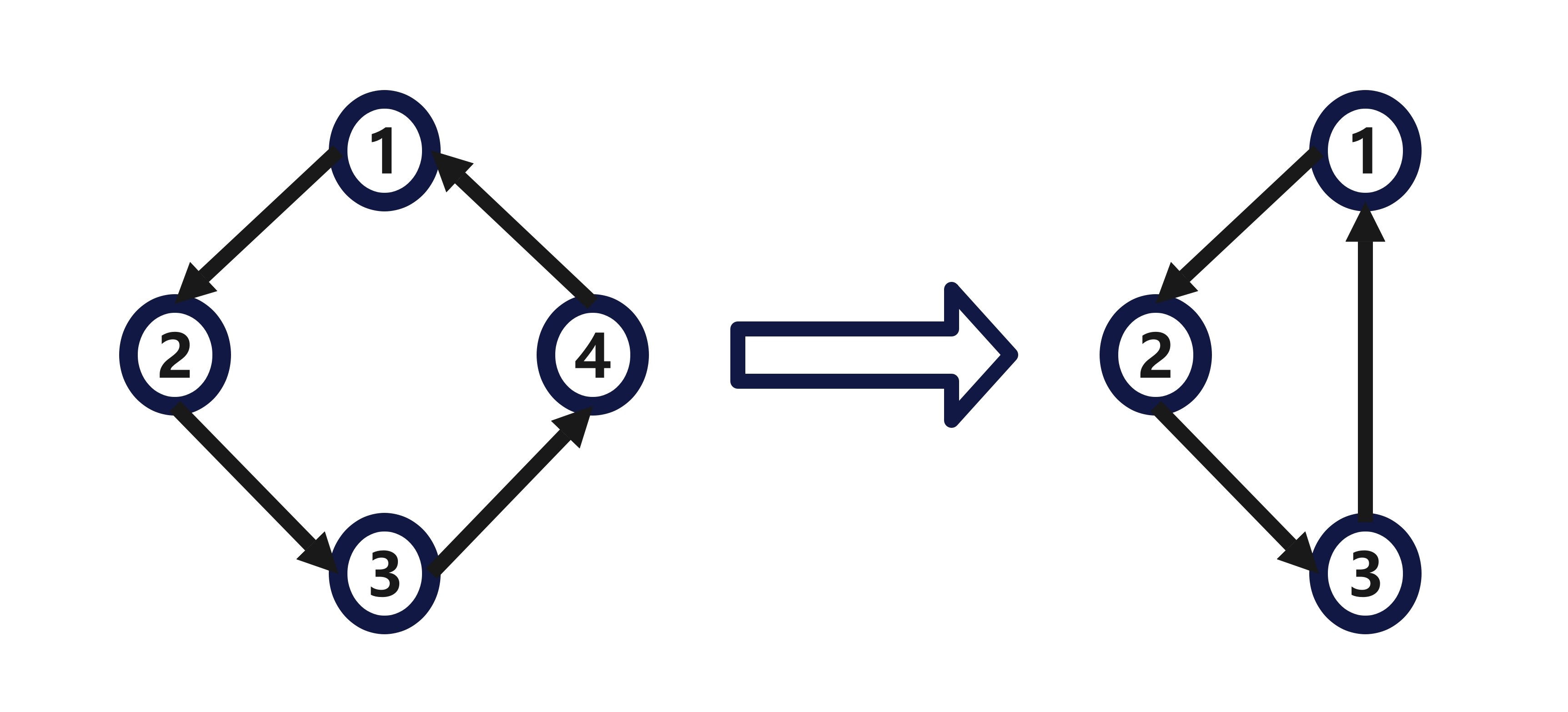}
\caption{Strongly-connected graphs corresponding to the weighted Laplacians before and after reduction (edge weights are omitted for simplicity)}
\label{SC-SC}
\end{figure}
\par Next we present theorems corresponding to \emph{Theorem} \ref{theorem-graph-to-Laplacian} but in the case of graphs being strongly-connected and quasi-strongly-connected.
\begin{theorem} \label{theoremSC-SC}
Suppose the corresponding graph $\mathcal{G}_d$ of the Laplacian matrix $\mathcal{L}=\mathcal{H}_{o}B\mathcal{H}_{}^T$ is strongly-connected. Then every Schur complement (if existing) of $\mathcal{L}$ can be written as $ \bar{\mathcal{H}}_{o } \bar{B}  \bar{\mathcal{H}}^{T} $, with $\bar{B}$ a positive definite diagonal matrix, and $\bar{\mathcal{H}}$ the incidence matrix of a strongly-connected directed graph $\bar{\mathcal{G}_d}$.
\end{theorem}
\begin{figure}[]
\centering
\includegraphics[width=0.45\textwidth]{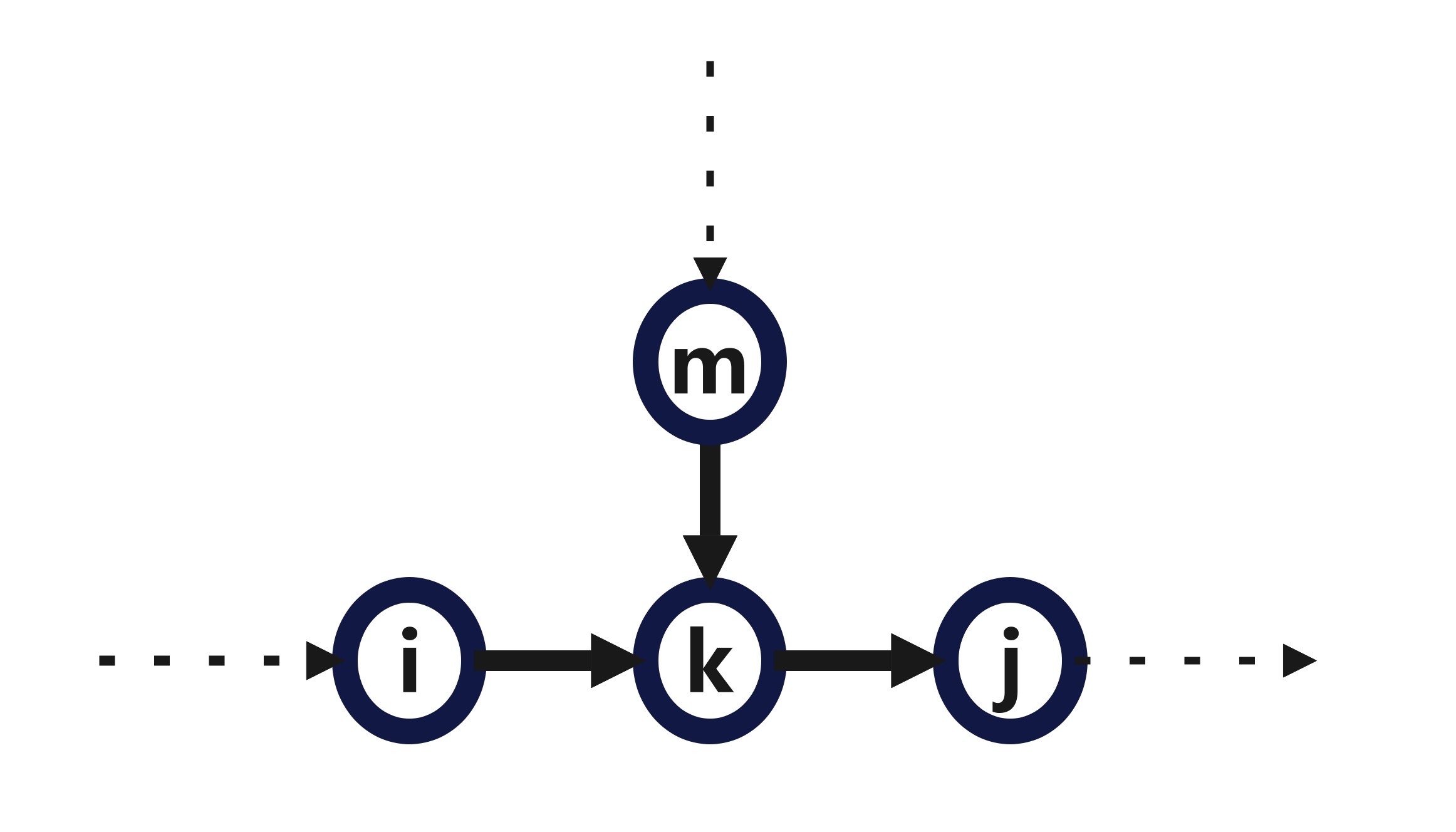}
\caption{Sub-graph consisting of the eliminated vertex $v_k$ and all of its adjacent vertices}
\label{iterative-subgraph}
\end{figure}
\begin{proof*}
From \emph{Remark} \ref{bbb-iterative} we know that \emph{block-by-block Kron reduction} is strictly equivalent to \emph{iterative Kron reduction} regarding reduction results. For the proof of \emph{Theorem} \ref{theoremSC-SC}, first, we consider our Kron reduction as a sequence of \emph{iterative Kron reduction}.
\par For each iterative step, we focus on the sub-graph consisting of the eliminated vertex $v_k$ and all of its adjacent vertices. For a better illustration without loss of generality, consider the sub-graph given in Fig. \ref{iterative-subgraph} as an example. Adjacent vertices of $v_k$ are $v_i$, $v_j$, and $v_m$. The sub-graph before reduction is associated with the adjacency matrix $\mathcal{A}_{sub}$ and the corresponding weighted Laplacian $\mathcal{L}_{red}$:
\begin{equation}
\mathcal{A}_{s u b}=\left[\begin{array}{cccc}
0 & b_{i k} & 0 & 0 \\
0 & 0 & b_{k j} & 0 \\
0 & 0 & 0 & 0 \\
0 & b_{m k} & 0 & 0
\end{array}\right], \mathcal{L}_{s u b}=\left[\begin{array}{cccc}
b_{i k} & -b_{i k} & 0 & 0 \\
0 & b_{k j} & -b_{k j} & 0 \\
0 & 0 & 0 & 0 \\
0 & -b_{m k} & 0 & b_{m k}
\end{array}\right]  \nonumber
\end{equation}

where non-zero entries in the adjacency matrix $\mathcal{A}_{sub}$ denote the edge weights on the edges from $v_i$ to $v_j$.
\par In the unreduced sub-graph in Fig. \ref{iterative-subgraph}, there are two nonzero walk products: $(\mathcal{A}_{sub}^2)_{ij}$ and $(\mathcal{A}_{sub}^2)_{mj}$ which are expressed in:
\begin{equation}
\begin{array}{l}
\left(\mathcal{A}_{s u b}^{2}\right)_{i j}=\left[\mathcal{A}_{s u b}\right]_{i k}\left[\mathcal{A}_{s u b}\right]_{k j}=b_{i k} b_{k j} \neq 0,  \\ 
\left(\mathcal{A}_{s u b}^{2}\right)_{m j}=\left[\mathcal{A}_{\text {sub }}\right]_{m k}\left[\mathcal{A}_{\text {sub }}\right]_{k j}=b_{m k} b_{k j} \neq 0.
\end{array}\nonumber
\end{equation}
\par By decomposing the weighted Laplacian $\mathcal{L}_{sub}$ as $\left[\begin{array}{cc}
\mathcal{L}_{s u b\{i, j, m\},\{i, j, m\}} & \mathcal{L}_{s u b\{i, j, m\},\{k\}} \\
\mathcal{L}_{s u b\{k\},\{i, j, m\}} & {\left[\mathcal{L}_{s u b}\right]_{k k}}
\end{array}\right]$ where $\mathcal{L}_{s u b\{i, j, m\},\{i, j, m\}}=\left[\begin{array}{ccc}
b_{i k} & 0 & 0 \\
0 & 0 & 0 \\
0 & 0 & b_{m k}
\end{array}\right]$, $\mathcal{L}_{s u b\{i, j, m\},\{k\}}=\left[\begin{array}{c}
-b_{i k} \\
0 \\
-b_{m k}
\end{array}\right]$, $\left[\mathcal{L}_{s u b}\right]_{k k}=b_{k j}$, and $\mathcal{L}_{s u b\{k\},\{i, j, m\}}=\left[\begin{array}{lll}
0 & -b_{k j} & 0
\end{array}\right]$, the iterative Kron reduction eliminating vertex $v_k$ can be formulated as the Schur complement of $\mathcal{L}_{sub}$ with respect to $[\mathcal{L}_{sub}]_{kk}$:

\begin{align}
\mathcal{L}_{s u b-r e d} & =\mathcal{L}_{s u b\{i, j, m\},\{i, j, m\}}-\mathcal{L}_{s u b\{i, j, m\},\{k\}}(\left[\mathcal{L}_{s u b}\right]_{k k})^{-1} \mathcal{L}_{s u b\{k\},\{i, j, m\}} \nonumber \\
& =\left[\begin{array}{ccc}
b_{i k} & -b_{i k} & 0 \\
0 & 0 & 0 \\
0 & -b_{m k} & b_{m k}
\end{array}\right]  \nonumber
\end{align}

\par The reduced adjacency matrix $\mathcal{A}_{sub-red}$ corresponding to the reduced weighted Laplacian is:

\begin{align}
\mathcal{A}_{\text {sub-red }}=\left[\begin{array}{ccc}
0 & b_{i k} & 0 \\
0 & 0 & 0 \\
0 & b_{m k} & 0
\end{array}\right].\nonumber
\end{align}

\par It is obvious that in the reduced adjacency matrix, there are two nonzero entries $[\mathcal{A}_{sub-red}]_{ij}$ and $[\mathcal{A}_{sub-red}]_{mj}$ which means that there exists a directed path from $v_i$ to $v_j$ and a directed path from $v_m$ to $v_j$. Non-zero walk products remain non-zero during each step of the iterative Kron reduction, hence non-zero walk products remain non-zero after the iterative Kron reduction. Therefore we can claim that non-zero walk products remain non-zero after the block-by-block Kron reduction. In other words, there exists a directed path from $v_i$ to $v_j$ in the reduced graph if there exists a directed path from $v_i$ to $v_j$ in the original graph.
\par In a strongly-connected graph there exists at least one directed path for every vertex to reach any other vertex in the graph. Hence there exists at least one directed path for every vertex to reach any other vertex in the reduced graph. Therefore, the reduced graph is again a strongly-connected graph.
\end{proof*}

\textbf{\emph{Example}} Consider the corresponding graph $\mathcal{G}_d$ of the Laplacian in Fig. \ref{SC-SC} (left). The weighted Laplacian $\mathcal{L}$ of the graph is:

\begin{align}
\mathcal{L}=\left[\begin{array}{cccc}
1 & -1 & 0 & 0 \\
0 & 1 & -1 & 0 \\
0 & 0 & 1 & -1 \\
-1 & 0 & 0 & 1
\end{array}\right]. \nonumber
\end{align}

The reduced Laplacian $\mathcal{L}_{red}$ and its corresponding incidence matrix $\bar{\mathcal{H}}$, variation of the incidence matrix $\bar{\mathcal{H}_o}$ and diagonal weighting matrix $\bar{B}$ are:

\begin{align}
\mathcal{L}_{\text {red }} & =\left[\begin{array}{ccc}
1 & -1 & 0 \\
0 & 1 & -1 \\
-1 & 0 & 1
\end{array}\right],  \bar{\mathcal{H}}=\left[\begin{array}{ccc}
1 & 0 & -1 \\
-1 & 1 & 0 \\
0 & -1 & 1
\end{array}\right], \nonumber \\
\bar{\mathcal{H}}_{o} & =\left[\begin{array}{lll}
1 & 0 & 0 \\
0 & 1 & 0 \\
0 & 0 & 1
\end{array}\right], \quad \quad  \bar{B}=\operatorname{diag}\{1,1,1\} \nonumber.
\end{align}

\par The reduced graph in Fig. \ref{SC-SC} (right) corresponding to the reduced incidence matrix $\bar{\mathcal{H}}$ is again a strongly-connected graph. This example illustrates that every Schur complement of the weighted Laplacian $\mathcal{L}$ corresponding to a strongly-connected graph $\mathcal{G}_d$ is again a weighted Laplacian matrix $\mathcal{L}_{red}$, which again corresponds to a strongly-connected graph. $\hfill \square$

\begin{theorem} \label{quais-quasi}
Suppose the corresponding graph $\mathcal{G}_d$ of the Laplacian matrix $\mathcal{L}=\mathcal{H}_{o}B\mathcal{H}_{}^T$ is quasi-strongly-connected. Then every Schur complement (if existing) of $\mathcal{L}$ can be written as $ \bar{\mathcal{H}}_{o } \bar{B}  \bar{\mathcal{H}}^{T} $, with $\bar{B}$ a positive definite diagonal matrix, and $\bar{\mathcal{H}}$ the incidence matrix of a quasi-strongly-connected directed graph $\bar{\mathcal{G}_d}$.
\end{theorem}
\begin{proof*}
We have proved that for any directed graph $\mathcal{G}_d$, there exists a directed path from $v_i$ to $v_j$ in the reduced graph if there exists a directed path from $v_i$ to $v_j$ in the original graph in the proof for \emph{Theorem} \ref{theoremSC-SC}. In a quasi-strongly-connected graph, there exists at least a \emph{source} vertex in the unreduced graph. Since \emph{source} vertices are \emph{boundary vertices} which will not be eliminated during Kron reduction, there exists a directed path for every other vertex (except for \emph{sink}) to start at \emph{sink} vertex and end at the very vertex in the reduced graph, hence concluding the proof for \emph{Theorem} \ref{quais-quasi}.
\end{proof*}
\par
\begin{figure}[]
\centering
\includegraphics[width=0.4\textwidth]{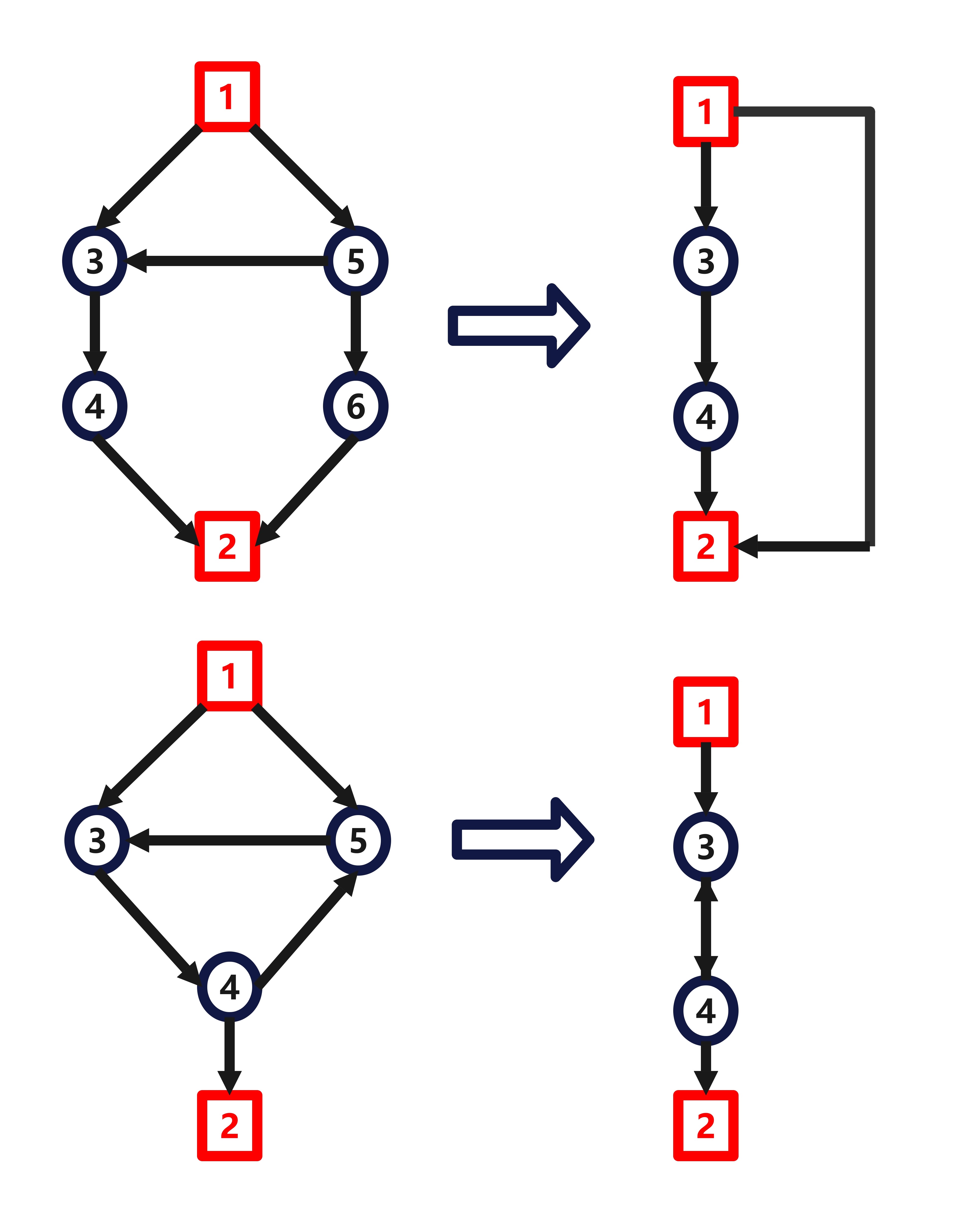}
\caption{Illustration of reduced graphs corresponding to Schur complements of two given Laplacians (Upper: $\mathcal{G}_d$ of $\mathcal{L}_1$ and $\mathcal{G}_{d\;red}$ of $\mathcal{L}_{1\;red}$. Bottom: $\mathcal{G}_d$ of $\mathcal{L}_2$ and $\mathcal{G}_{d\;red}$ of $\mathcal{L}_{2\;red}$). Edge weights are omitted for simplicity.}
\label{qsc_sink_source_acyclic_cyclic}
\end{figure}

 \textbf{\emph{Example}} Consider the corresponding quasi-strongly-connected graphs in Fig. \ref{qsc_sink_source_acyclic_cyclic} of two given Laplacians $\mathcal{L}_1$ and $\mathcal{L}_2$: 
\begin{small}

\begin{align}
\mathcal{L}_{1}&=\left[\begin{array}{cccccc}
2 & 0 & -1 & 0 & -1 & 0 \\
0 & 0 & 0 & 0 & 0 & 0 \\
0 & 0 & 1 & -1 & 0 & 0 \\
0 & -1 & 0 & 1 & 0 & 0 \\
0 & 0 & -1 & 0 & 2 & -1 \\
0 & -1 & 0 & 0 & 0 & 1
\end{array}\right], \quad \nonumber \\ \mathcal{L}_{2}&=\left[\begin{array}{ccccc}
2 & 0 & -1 & 0 & -1 \\
0 & 0 & 0 & 0 & 0 \\
0 & 0 & 1 & -1 & 0 \\
0 & -1 & 0 & 2 & -1 \\
0 & 0 & -1 & 0 & 1
\end{array}\right]. \nonumber
\end{align}

\end{small}
\par The reduced Laplacians $\mathcal{L}_{1\;red}$ and $\mathcal{L}_{2\;red}$, the corresponding incidence matrices $\bar{\mathcal{H}}_1$ and $\bar{\mathcal{H}}_2$, their variations $\bar{\mathcal{H}}_{1o}$ and $\bar{\mathcal{H}}_{2o}$ and corresponding diagonal weighting matrices $\bar{B}_1$ and $\bar{B}_2$ are: 
\begin{footnotesize}
\begin{align}
\begin{array}{ccc}
\mathcal{L}_{1 \text { red }}=\left[\begin{array}{cccc}
2 & -0.5 & -1.5 & 0 \\
0 & 0 & 0 & 0 \\
0 & 0 & 1 & -1 \\
0 & -1 & 0 & 1
\end{array}\right], & \mathcal{L}_{2 \text { red }}=\left[\begin{array}{cccc}
2 & 0 & -2 & 0 \\
0 & 0 & 0 & 0 \\
0 & 0 & 1 & -1 \\
0 & -1 & -1 & 2
\end{array}\right], \\
\\
\overline{\mathcal{H}}_{1}=\left[\begin{array}{cccc}
1 & 0 & 0 & 1 \\
0 & 0 & -1 & -1 \\
-1 & 1 & 0 & 0 \\
0 & -1 & 1 & 0
\end{array}\right], & \overline{\mathcal{H}}_{2}=\left[\begin{array}{cccc}
1 & 0 & 0 & 0 \\
0 & 0 & 0 & -1 \\
-1 & 1 & -1 & 0 \\
0 & -1 & 1 & 1
\end{array}\right], \\
\\
\overline{\mathcal{H}}_{1 o}=\left[\begin{array}{cccc}
1 & 0 & 0 & 1 \\
0 & 0 & 0 & 0 \\
0 & 1 & 0 & 0 \\
0 & 0 & 1 & 0
\end{array}\right], & \overline{\mathcal{H}}_{2 o}=\left[\begin{array}{cccc}
1 & 0 & 0 & 0 \\
0 & 0 & 0 & 0 \\
0 & 1 & 0 & 0 \\
0 & 0 & 1 & 1
\end{array}\right], \\
\\
\bar{B}_{1}=\operatorname{diag}\{1.5,1,1,0.5\}, & \bar{B}_{2}=\operatorname{diag}\{2,1,1,1\} . \nonumber
\end{array}
\end{align} 
\end{footnotesize}
\par The reduced graphs (Fig. \ref{qsc_sink_source_acyclic_cyclic}, right) corresponding to the reduced Laplacians are still quasi-strongly-connected. This example illustrates that every existing Schur complement of the weighted Laplacian $\mathcal{L}$ corresponding to a quasi-strongly-connected graph $\mathcal{G}_d$ is again a weighted Laplacian matrix $\mathcal{L}_{red}$, which again corresponds to a quasi-strongly-connected graph. $\hfill \square$

\section{\large Kron reduction to power flow networks} \label{section-main-results-kron}

\par In this section, we present the graph-theoretic analysis of the Kron reduction process on DC power flow networks.
\subsection{Vertex classification} \label{subsection-vertex-class}

\begin{figure}[]
\centering
\includegraphics[width=0.25 \textwidth]{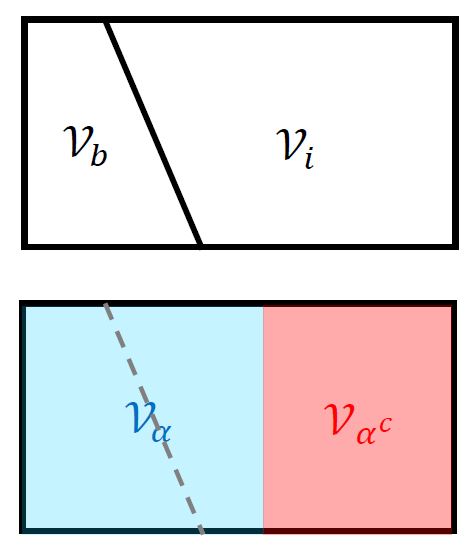}
\caption{Illustration of vertex classification, $\mathcal{V}_b\cap\mathcal{V}_i=\mathcal{V}$ (upper), $\mathcal{V}_\alpha\cap\mathcal{V}_{\alpha^c}=\mathcal{V},\mathcal{V}_{\alpha^c}\subseteq \mathcal{V}_i$ (bottom)}
\label{verticeclass}
\end{figure}   

\par In this subsection we will identify a set of vertices that are actually eliminated and a set of vertices that are actually retained during the Kron reduction process. First, recall that for a given graph $\mathcal{G}_d$, we identified a subset $\mathcal{V}_{b} \subset \mathcal{V}$ as \emph{boundary vertices} and a subset $\mathcal{V}_i=\mathcal{V}\setminus\mathcal{V}_{b}$ as \emph{interior vertices} in Section \ref{subsection_weighted Laplacian matrix}. Boundary vertices are vertices that \textbf{cannot} be eliminated. Interior vertices are vertices that \textbf{can} be eliminated. Although all \emph{interior vertices} \textbf{can} be eliminated, there are times during Kron reduction when some of the interior vertices are to be retained. Hereby we further identify a subset termed \emph{eliminated vertices} $\mathcal{V}_{\alpha^c} \subseteq \mathcal{V}_i$ being the vertices that are \textbf{actually eliminated} during the Kron reduction process, and the subset termed \emph{retained vertices} $\mathcal{V}_{\alpha}=\mathcal{V}\setminus \mathcal{V}_{\alpha^c}$ being the vertices that are \textbf{actually retained} during reduction. See Fig.~\ref{verticeclass} for a diagrammatic illustration of vertex classification.

\par Recall the power-angle equation (\ref{Pv=Ltheta}) in Section \ref{subsection_weighted Laplacian matrix}. Decompose $P_v$ as $\left[\begin{array}{c}
P_{v \alpha} \\
P_{v \alpha^{c}}
\end{array}\right]$ with $P_{v \alpha}$ corresponding to active power extractions at retained vertices and $P_{v \alpha^{c}}$ corresponding to active power extractions at eliminated vertices. Decompose $\theta$ as $\left[\begin{array}{c}
\theta_{\alpha} \\
\theta_{\alpha^{c}}
\end{array}\right]$ with $\theta_{\alpha}$ corresponding to angles at retained vertices and $\theta_{\alpha^c}$ angles at eliminated vertices. Further decompose $\mathcal{L}$ as $\left[\begin{array}{ll}
\mathcal{L}_{\alpha \alpha} & \mathcal{L}_{\alpha \alpha^{c}} \\
\mathcal{L}_{\alpha^{c} \alpha} & \mathcal{L}_{\alpha^{c} \alpha^{c}}
\end{array}\right]$ with subblocks being composed of columns and rows corresponding retained and eliminated vertices respectively. Then (\ref{Pv=Ltheta}) can be partitioned as

\begin{align}
\left[\begin{array}{c}
P_{v \alpha} \\
P_{v \alpha^{c}}
\end{array}\right]=\left[\begin{array}{ll}
\mathcal{L}_{\alpha \alpha} & \mathcal{L}_{\alpha \alpha^{c}} \\
\mathcal{L}_{\alpha^{c} \alpha} & \mathcal{L}_{\alpha^{c} \alpha^{c}}
\end{array}\right]\left[\begin{array}{c}
\theta_{\alpha} \\
\theta_{\alpha^{c}}
\end{array}\right]. \label{P=Lthetapartition}
\end{align}

Gaussian elimination of \emph{eliminated angles} $\theta_{\alpha^c}$ in (\ref{P=Lthetapartition}) gives a reduced network with \emph{retained vertices} obeying the reduced power flow equations

\begin{align}
P_{v\alpha}+\mathcal{L}_{ac}P_{v{\alpha^c}}=\mathcal{L}_{red}\theta_{\alpha} \label{kronreducedpowerflow}
\end{align}

\noindent where the reduced Laplacian matrix is given by the Schur complement of $\mathcal{L}$ with respect to \emph{retained vertices} $\mathcal{V}_\alpha$, that is $\mathcal{L}_{red}=\mathcal{L}_{\alpha \alpha}-\mathcal{L}_{\alpha \alpha^{c}}\mathcal{L}_{\alpha^{c} \alpha^{c}}^{-1}\mathcal{L}_{\alpha^{c} \alpha}$ and the accompanying matrix $\mathcal{L}_{ac}=-\mathcal{L}_{\alpha \alpha^{c}}\mathcal{L}_{\alpha^{c} \alpha^{c}}^{-1}$ maps \emph{eliminated active power extractions} $P_{v{\alpha^c}}$ to \emph{retained active power extractions} $P_{vred}=P_{v\alpha}+\mathcal{L}_{ac}P_{v{\alpha^c}}$ in the reduced network.

\subsection{Kron reduction to power flow networks} 
\par Following the identification of \emph{retained vertices} and \emph{eliminated vertices} in the last section, we formally give the definition of Kron reduction to power flow networks.
\begin{figure}[]
\centering
\includegraphics[width=0.45 \textwidth]{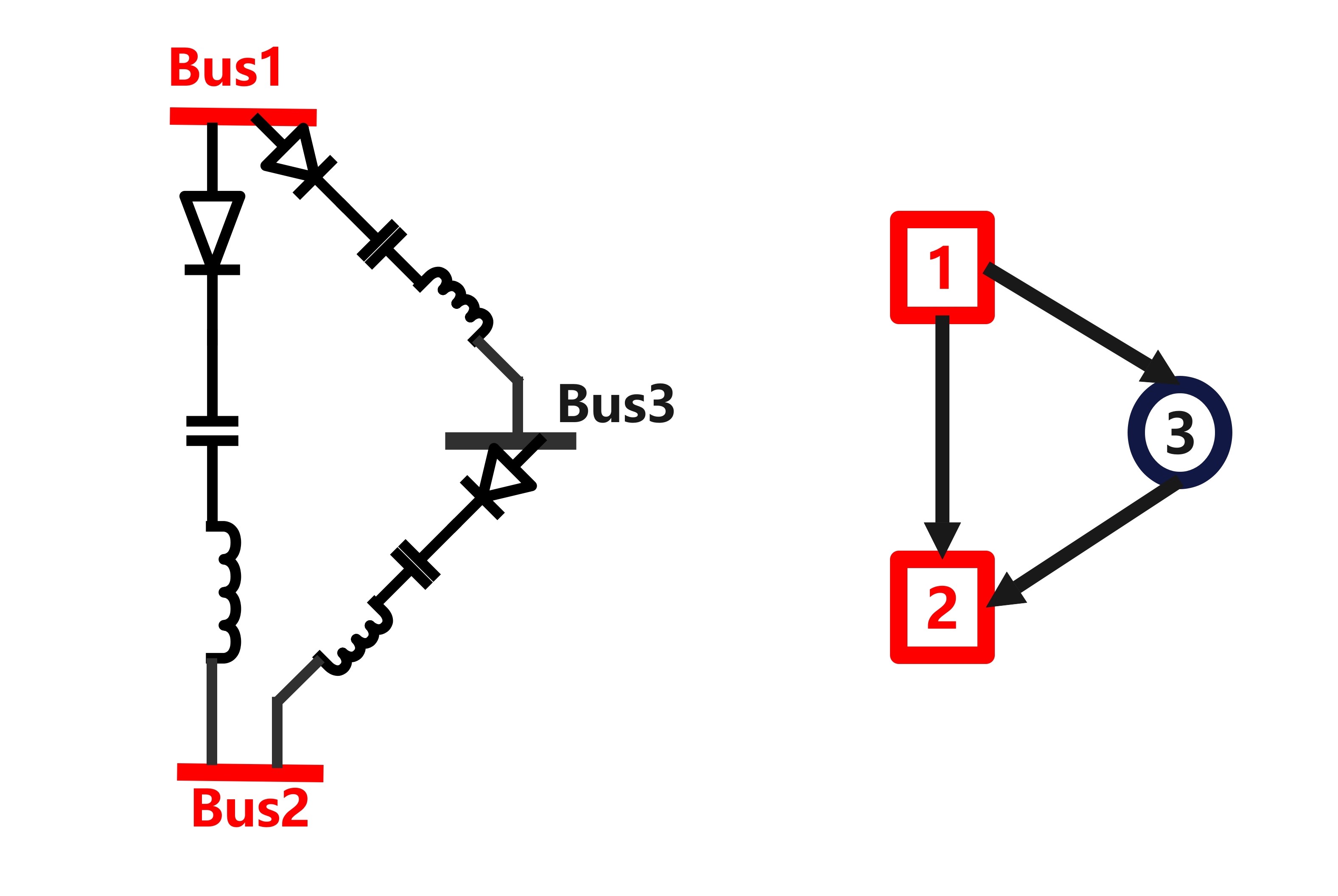}
\caption{IEEE-3 test feeder and its directed graph representation (boundary vertices marked in red)}
\label{IEEE3}
\end{figure} 
\begin{definition}[{Kron reduction to power flow networks}] \label{Def_Kronreduction}
Consider a power flow network corresponding to the graph representation $\mathcal{G}_d=(\mathcal{V},\varepsilon_d,\mathcal{H}, B)$. Let $\mathcal{L}\in \mathbb{R}^{|\mathcal{V}|\times |\mathcal{V}|}$ denote the weighted Laplacian matrix: $\mathcal{H}_{o}B\mathcal{H}^T$ of the graph. Let $\mathcal{V}_{\alpha}\subset\mathcal{V}$, \textbf{the} \emph{retained vertices} be a \textbf{proper} subset of vertices with $|\mathcal{V}_{\alpha}|\ge2$. (`\textbf{Proper}' means that \emph{boundary vertices} are always included in $\mathcal{V}_{\alpha}$ following the classification in Section \ref{subsection-vertex-class}.) Then the $|\mathcal{V}_{\alpha}|\times|\mathcal{V}_{\alpha}|$ dimensional Kron reduced matrix $\mathcal{L}_{red}$ is defined by the Schur complement of $\mathcal{L}$ with respect to the sub-matrix consisting of rows and columns corresponding to \emph{retained vertices}:

\begin{align}
\mathcal{L}_{red}=\mathcal{L}_{\alpha \alpha}-\mathcal{L}_{\alpha \alpha^{c}}\mathcal{L}_{\alpha^{c} \alpha^{c}}^{-1}\mathcal{L}_{\alpha^{c} \alpha}, \label{DefinitionKronreductiontoPF}
\end{align}

\noindent which gives the reduced power flow network with the reduced graph representation i.e., $\bar{\mathcal{G}}_d=(\mathcal{V_{\alpha}},\bar{\varepsilon}_d,\bar{\mathcal{H}},\bar{\mathcal{B}})$, with $\mathcal{L}_{red}=\mathcal{L}_{\alpha \alpha}-\mathcal{L}_{\alpha \alpha^{c}}\mathcal{L}_{\alpha^{c} \alpha^{c}}^{-1}\mathcal{L}_{\alpha^{c} \alpha}$.
\end{definition}

\begin{remark}
In most cases (most IEEE test feeders) power flow networks are neither strongly-connected nor quasi-strongly-connected. However, there are several cases when power flow networks are relatively simple and are quasi-strongly-connected. See the example of an IEEE-3 test feeder, in Fig. \ref{IEEE3}. Vertex $1$ is the \emph{root vertex} of this quasi-strongly-connected graph.
\end{remark}

\par Next, we discuss sufficient conditions for the existence of Kron reduction to power flow networks.

\begin{lemma}[{Existence of Kron reduction to power flow networks with quasi-strongly-connected graph representations}]
Consider a power flow network corresponding to the quasi-strongly-connected graph representation $\mathcal{G}_d=(\mathcal{V},\varepsilon_d,\mathcal{H},B)$ with the weighted Laplacian $\mathcal{L}=\mathcal{H}_{o}B\mathcal{H}^T$. Let $\mathcal{V}_{\alpha}\subset\mathcal{V}$, the \emph{retained vertices} be a proper subset of vertices with $|\mathcal{V}_{\alpha}|\ge2$. Then Kron reduction always exists for this network.
\end{lemma}
\begin{proof*}
For a given power flow network corresponding to the quasi-strongly-connected graph $\mathcal{G}_d$ with the weighted Laplacian $\mathcal{L}$, since $\mathcal{V}_{sink} \in \mathcal{V}_b \subset \mathcal{V}_{\alpha}$, Schur complements of $\mathcal{L}$ with respect to sub-matrices consisting of rows and columns corresponding to $\mathcal{V}_{\alpha}$ always exist by referring to \emph{Lemma} \ref{QSC-GRAPH-LAPLACIAN}.3. Therefore, Kron reduction always exists for this network. Laplacian matrix of the reduced network is given by (\ref{DefinitionKronreductiontoPF}).
\end{proof*}

\begin{figure}[]
\centering
\includegraphics[width=0.6 \textwidth]{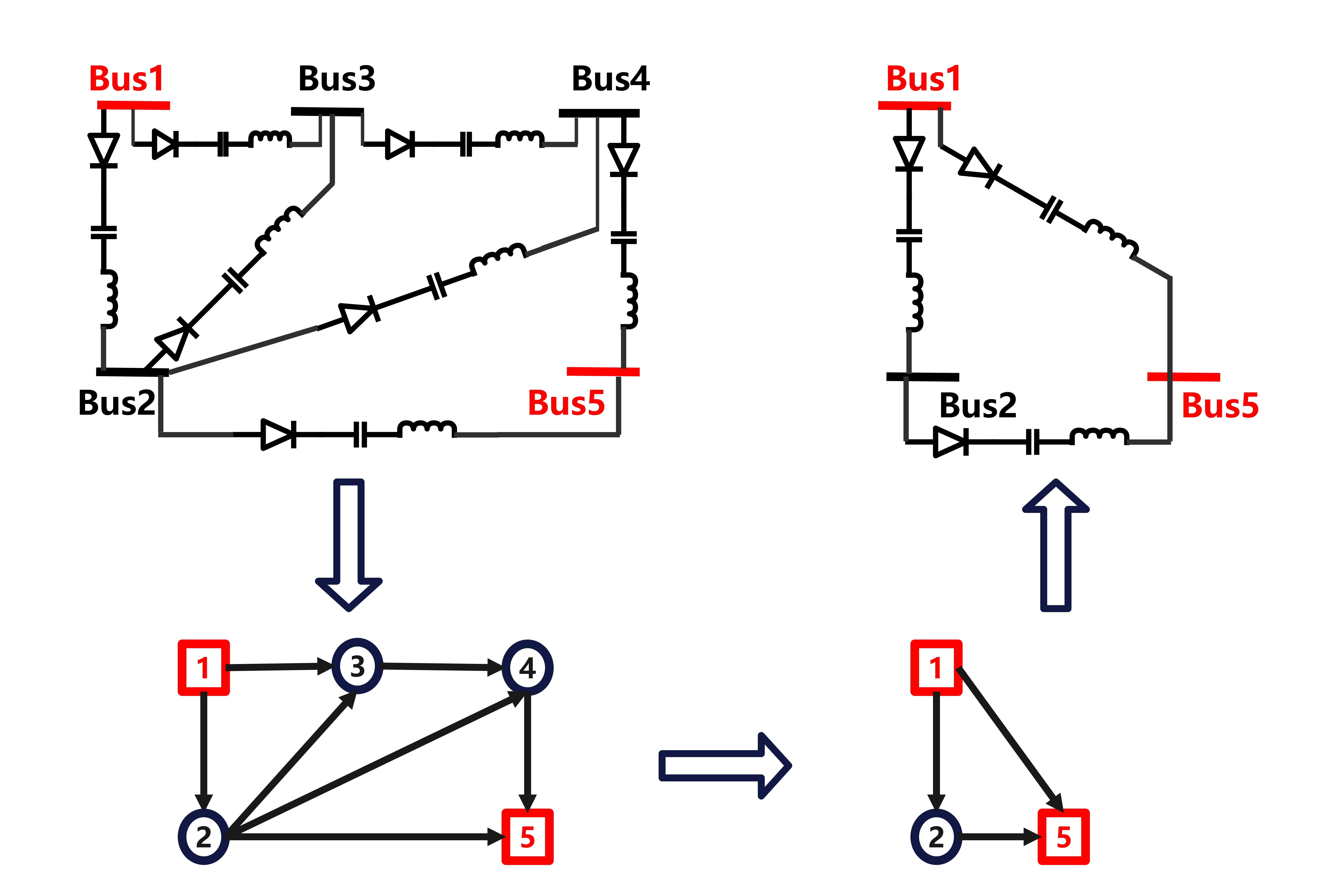}
\caption{IEEE-5 test feeder, its directed graph representation (boundary vertices marked in red), the reduced graph representation, and the restored reduced network}
\label{IEEE5-reduced}
\end{figure} 
\textbf{\emph{Example}} For the quasi-strongly-connected corresponding graph representation of an IEEE-5 test feeder in Fig. \ref{IEEE5-reduced}, vertex $1$ and vertex $5$ are boundary vertices, which are included in $\mathcal{V}_{\alpha}$. Kron reduction of this network eliminates vertex $2$; see Fig. \ref{IEEE5-reduced}. Assume all edge susceptances are 1. The graph of the reduced network is quasi-strongly-connected, which conforms to \emph{Theorem} \ref{quais-quasi}. The weighted Laplacian $\mathcal{L}$ for the original network and the weighted Laplacian $\mathcal{L}_{red}$ for the reduced network are:

\begin{align}
\mathcal{L}=\left[\begin{array}{ccccc}
2 & -1 & -1 & 0 & 0 \\
0 & 3 & -1 & -1 & -1 \\
0 & 0 & 1 & -1 & 0 \\
0 & 0 & 0 & 1 & -1 \\
0 & 0 & 0 & 0 & 0
\end{array}\right], \quad \mathcal{L}_{\text {red }}=\left[\begin{array}{ccc}
2 & -1 & -1 \\
0 & 3 & -3 \\
0 & 0 & 0
\end{array}\right]. \nonumber
\end{align}

\par This example illustrates that Kron reduction exists for a lossless DC power flow network that corresponds to a quasi-strongly-connected graph. $\hfill \square$

\begin{figure}[]
\centering
\includegraphics[width=0.8 \textwidth]{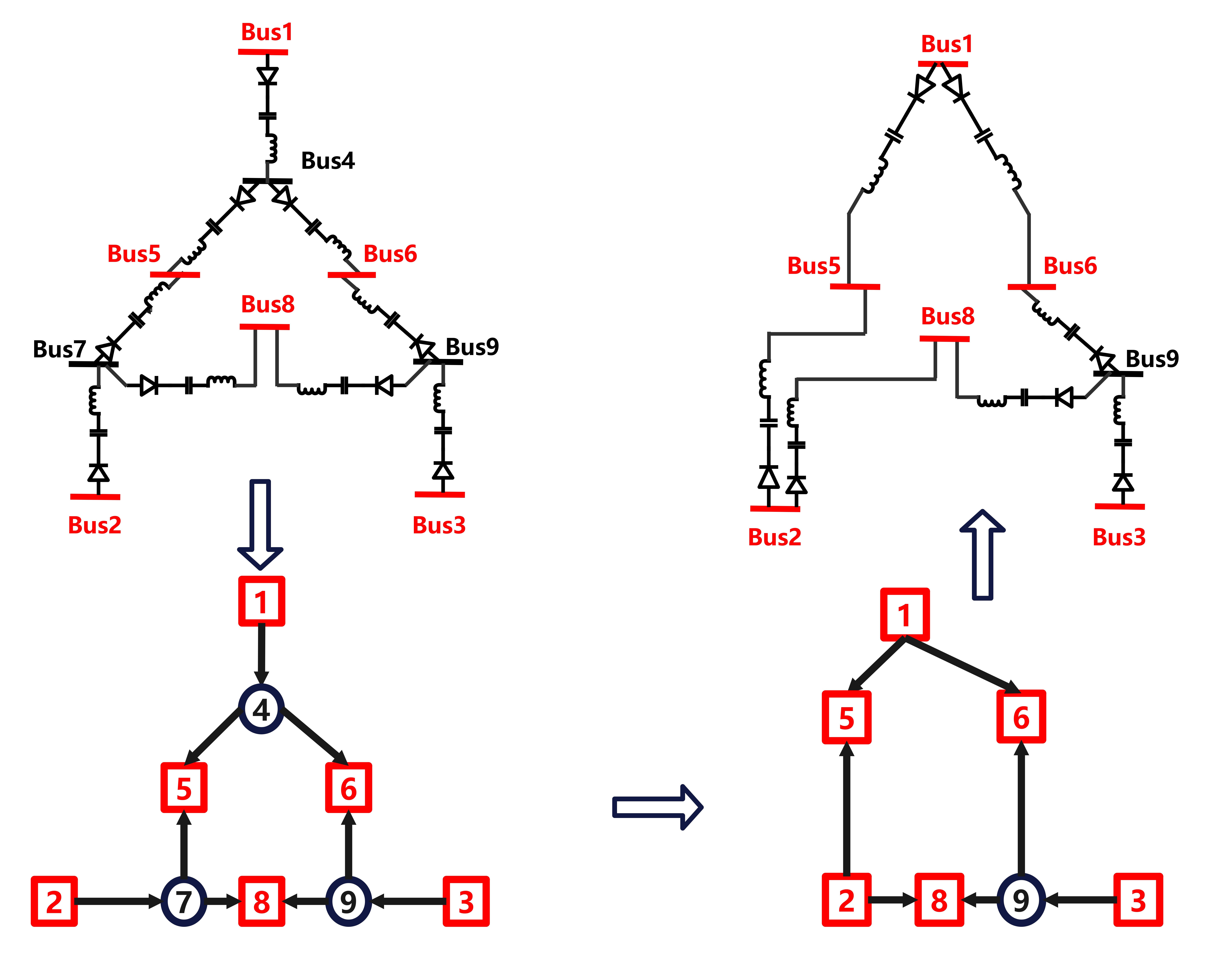}
\caption{IEEE-9 test feeder, its directed graph representation (boundary vertices marked in red), the reduced graph representation, and the restored reduced network}
\label{IEEE9-reduced}
\end{figure}

\begin{lemma}[{Existence of Kron reduction to generalized power flow networks}] \label{lemmageneralizedPF}
Consider a generalized power flow network with the graph representation $\mathcal{G}_d=(\mathcal{V},\varepsilon_d,\mathcal{H}, B)$ that consists of \emph{sink} vertices and \emph{source} vertices. Let $\mathcal{V}_{\alpha}\subset\mathcal{V}$, the \emph{retained vertices} be a proper subset of vertices with $|\mathcal{V}_{\alpha}|\ge2$. Then Kron reduction always exists for this network. \label{lemma-kron-tree}
\end{lemma}
\begin{proof*}
For a given power flow network of which the graph is not quasi-strongly-connected but still consists of \emph{sink} vertices and \emph{source} vertices, since $\mathcal{V}_{sink} \in \mathcal{V}_b \subset \mathcal{V}_{\alpha}$, there exists a directed path in $\mathcal{G}_d$ starting at any vertex in $\mathcal{V}_{{\alpha}^c}$ and ending at a \emph{sink} vertex in $\mathcal{V}_{\alpha}$. Therefore $\mathcal{V}_{\alpha}$ is always a \emph{reachable subset} of $\mathcal{G}_d$ for $\mathcal{V}_{{\alpha}^c}$. By referring to \emph{Lemma} \ref{lemma-sugiyama}, we can declare Schur complements of $\mathcal{L}$ with respect to sub-matrices consisting of rows and columns corresponding to $\mathcal{V}_{\alpha}$ always exist. The proof for  \emph{Lemma} \ref{lemmageneralizedPF} is concluded.
\end{proof*}
 \textbf{\emph{Example}} For the corresponding graph representation of an IEEE-9 test feeder in Fig. \ref{IEEE9-reduced}, vertices $1,2,3,5,6,8$ are boundary vertices, which are included in $\mathcal{V}_{\alpha}$. The graph representation of this network is not quasi-strongly-connected. Kron reduction of this network eliminates vertex $4,7$; see Fig. \ref{IEEE9-reduced}. Assume all edge susceptances are 1. The weighted Laplacian $\mathcal{L}$ for the original network and the weighted Laplacian $\mathcal{L}_{red}$ for the reduced network are:

\begin{align}
\begin{array}{c}
\mathcal{L}=\left[\begin{array}{ccccccccc}
1 & 0 & 0 & -1 & 0 & 0 & 0 & 0 & 0 \\
0 & 1 & 0 & 0 & 0 & 0 & -1 & 0 & 0 \\
0 & 0 & 1 & 0 & 0 & 0 & 0 & 0 & -1 \\
0 & 0 & 0 & 2 & -1 & -1 & 0 & 0 & 0 \\
0 & 0 & 0 & 0 & 0 & 0 & 0 & 0 & 0 \\
0 & 0 & 0 & 0 & 0 & 0 & 0 & 0 & 0 \\
0 & 0 & 0 & 0 & -1 & 0 & 2 & -1 & 0 \\
0 & 0 & 0 & 0 & 0 & 0 & 0 & 0 & 0 \\
0 & 0 & 0 & 0 & 0 & -1 & 0 & -1 & 2
\end{array}\right], \nonumber \quad \mathcal{L}_{\text {red }}=\left[\begin{array}{ccccccc}
1 & 0 & 0 & -0.5 & -0.5 & 0 & 0 \\
0 & 1 & 0 & -0.5 & 0 & -0.5 & 0 \\
0 & 0 & 1 & 0 & 0 & 0 & -1 \\
0 & 0 & 0 & 0 & 0 & 0 & 0 \\
0 & 0 & 0 & 0 & 0 & 0 & 0 \\
0 & 0 & 0 & 0 & 0 & 0 & 0 \\
0 & 0 & 0 & 0 & -1 & -1 & 2
\end{array}\right]. \nonumber
\end{array}
\end{align}
\\

\par This example illustrates that Kron reduction exists for a lossless DC power flow network that corresponds to a weighted directed graph consisting of \emph{sink} and \emph{source} vertices. $\hfill \square$

\subsection{Input-output behaviors of lossless DC power flow networks}
\par In this subsection, we present how the weighted Laplacian matrix $\mathcal{L}$ and its Kron-reduced form $\mathcal{L}_{red}$ function as I/O mappings for a lossless power flow system.
\begin{theorem} \label{theorem-IO-of-lossless-PF}
Consider a lossless DC power flow network with the graph representation $\mathcal{G}_d=(\mathcal{V},\varepsilon_d,\mathcal{H}, B)$ of which \emph{boundary vertices} consist of both \emph{sink} and \emph{source vertices}. The corresponding weighted Laplacian is 
$\mathcal{L}=\mathcal{H}_{o}B\mathcal{H}^T$. Then

\begin{enumerate}
\item The Laplacian $\mathcal{L}$ maps vertex angle vector $\theta$ (input) to vertex power extraction vector $P_v$ (output).
\item For any retained vertex angle vector $\theta_{\alpha}$, there exists a unique $\mathcal{L}_{red}$ such that (\ref{kronreducedpowerflow}) is satisfied.
\item To any weighted directed Laplacian matrix $\mathcal{L}_{red}$ there corresponds a lossless DC power flow network of which the input-output behavior is given by the linear map:

\begin{align}
P_{vred}=\mathcal{L}_{red} \theta_{\alpha}.  \label{reduced-power-expression}
\end{align}

\end{enumerate}

\end{theorem}
\begin{proof*}
\begin{enumerate}
\item For the proof of \emph{Theorem} \ref{theorem-IO-of-lossless-PF}.1, we aim to show that $\mathcal{L}$ indeed functions as a mapping from $\theta$ to $P_v$. Recall the expression (\ref{Pv=Ltheta}), the statement in \emph{Theorem} \ref{theorem-IO-of-lossless-PF}.1 is evident.
\item For the proof of \emph{Theorem} \ref{theorem-IO-of-lossless-PF}.2, we first aim to prove that for any retained vertex subset, the Schur complement of the reduction always exists. Consider the directed graph representation $\mathcal{G}_d$ corresponding to the given lossless DC power flow network. Since both \emph{sink} and \emph{source} vertices are boundary vertices, they are not eliminated. Therefore for any vertex $v_i\in \mathcal{V}_{{\alpha}^c}$, there exists a directed path in $\mathcal{G}_d$ starting at the very vertex and ending at the \emph{sink} vertex. Hence $\mathcal{V}_{{\alpha}}$ is a reachable subset for $\mathcal{V}_{{\alpha}^c}$. By recalling \emph{Lemma} \ref{lemma-sugiyama}, we can declare that for any retained vertex subset, the Schur complement of the reduction always exists, which also means that $\mathcal{L}_{\alpha^{c} \alpha^{c}}$ is non-singular and $\mathcal{L}_{red}=\mathcal{L}_{\alpha \alpha}-\mathcal{L}_{\alpha \alpha^{c}}\mathcal{L}_{\alpha^{c} \alpha^{c}}^{-1}\mathcal{L}_{\alpha^{c} \alpha}$ exists for (\ref{kronreducedpowerflow}). Hence we conclude the proof for \emph{Theorem} \ref{theorem-IO-of-lossless-PF}.2.
\item Following the proof for \emph{Theorem} \ref{theorem-IO-of-lossless-PF}.2, the left hand side of (\ref{kronreducedpowerflow}) is precisely the expression for $P_{red}$. Hence we have $P_{red}=\mathcal{L}_{red}\theta_{\alpha}$. According to our proof for \emph{Theorem} \ref{theorem-graph-to-Laplacian}, the reduced Laplacian $\mathcal{L}_{red}$ corresponds to a weighted directed graph, from which the reduced lossless DC power flow network can be restored.
\end{enumerate}
\end{proof*}
\begin{remark}
An example illustrating \emph{Theorem} \ref{theorem-IO-of-lossless-PF} will be given in Section \ref{section-numericalresults}. In Section \ref{section-mainresults} we presented several important properties of the weighted Laplacians of different types of directed graphs. In Section \ref{section-main-results-kron} we presented the methodology of using directed graphs to model lossless power flow networks and the physical interpretation of the weighted Laplacian. This work can be viewed as an extension of the work in \cite{van2010characterization} by A. van der Schaft.
\end{remark}

\section{\large Numerical results} \label{section-numericalresults}
\begin{figure}[]
\centering
\includegraphics[width=0.7\textwidth]{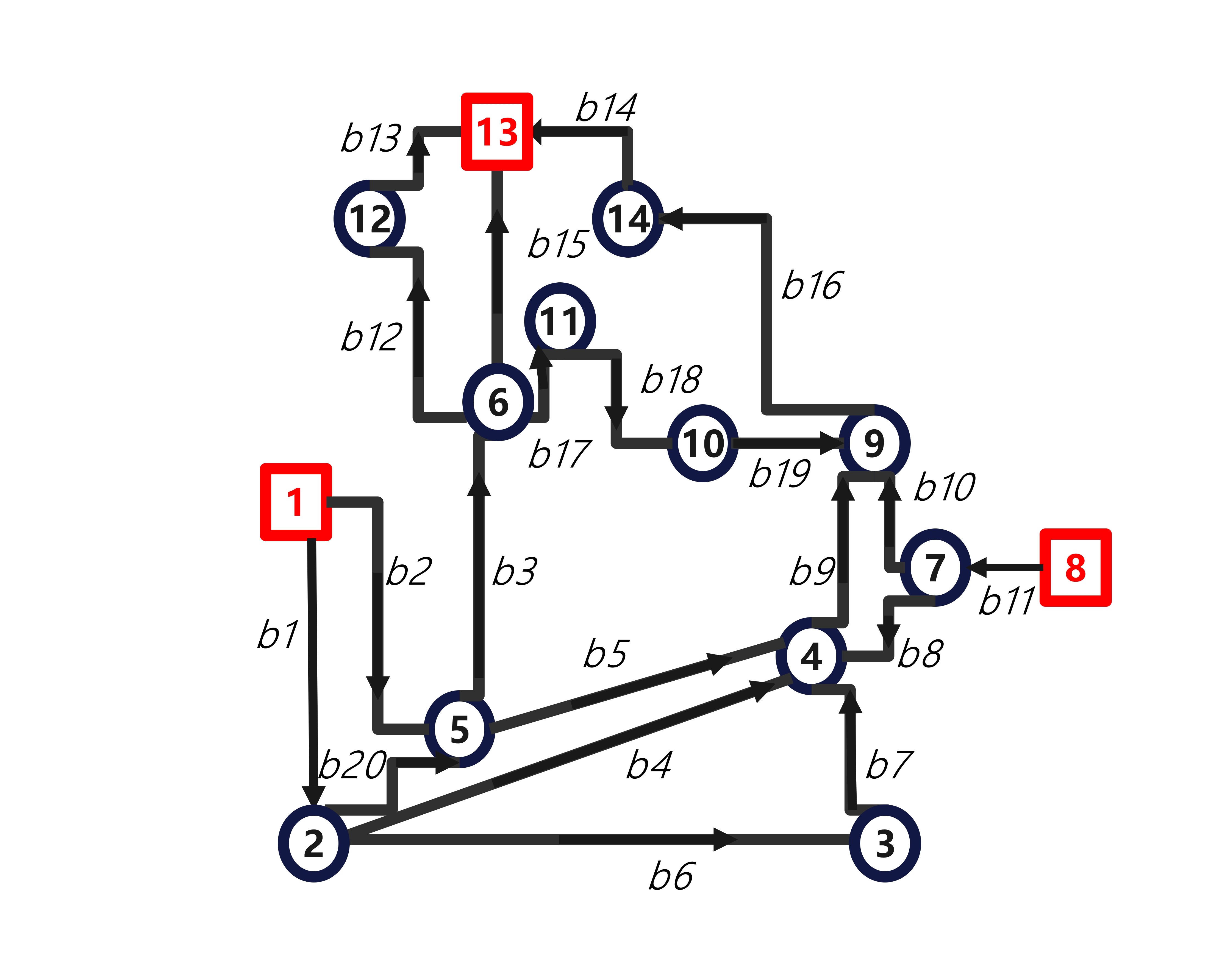}
\caption{Graph representation of IEEE-14 test feeder, boundary vertices marked in red squares}
\label{IEEE14_SINK1_SOURCE2}
\end{figure}

\par In this section, the IEEE-14 test feeder will be used as a detailed example for numerical testing; see the weighted graph representation of the IEEE-14 power flow network in Fig. \ref{IEEE14_SINK1_SOURCE2}. A two-stage reduction process will be adopted. During the first stage, boundary vertices and vertices that are connected to boundary vertices via an edge are retained. During Stage \uppercase\expandafter{\romannumeral2} where the reduction is performed on the reduced result of Stage \uppercase\expandafter{\romannumeral1}, all interior vertices are to be eliminated. Testing on a modified IEEE RTS-96 test system will also be presented in order to show the scalability of the proposed reduction method. 

\subsection{IEEE-14 test feeder}
\subsubsection{Reduction process}
\begin{figure}[]
\centering
\includegraphics[width=0.35\textwidth]{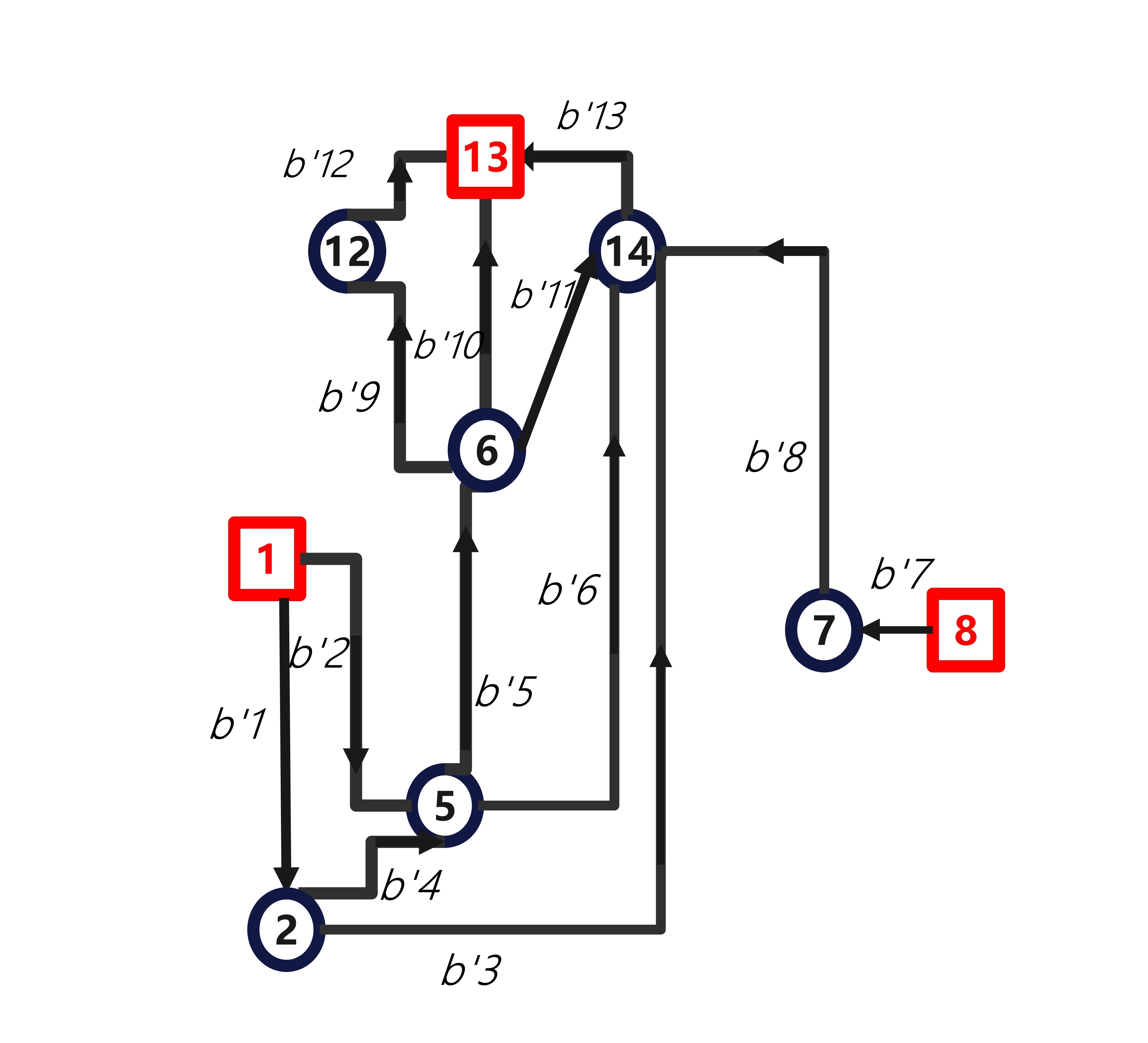}
\caption{Reduced IEEE-14 test feeder with vertices 3,4,9,10,11 eliminated }
\label{SINK1_SOURCE2_REDUC_1}
\end{figure}

\begin{figure}[]
\centering
\includegraphics[width=0.3\textwidth]{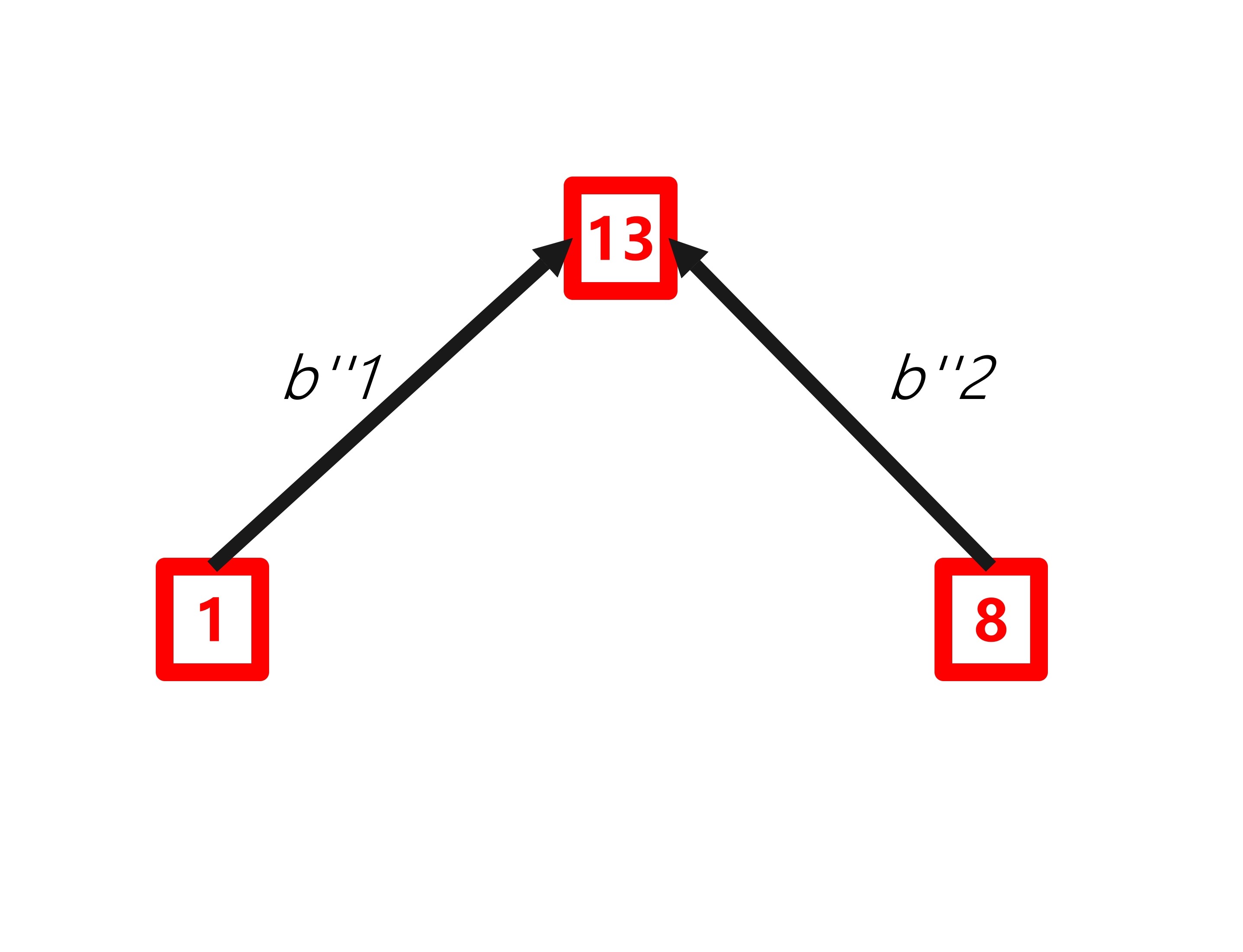}
\caption{Reduced IEEE-14 test feeder with all interior vertices eliminated}
\label{SINK1_SOURCE2_REDUC_2}
\end{figure}

The reduction process is detailed as follows:
\begin{enumerate}
\item Vertex classification: Each bus of the IEEE-14 network corresponds to a vertex of the graph. Buses that are connected to generators (outside the IEEE-14 network) correspond to \emph{source vertices}. Relatively, buses that are 
connected to loadings (outside the IEEE-14 network) correspond to \emph{sink vertices}. Buses that are connected to both generators and loadings (outside the IEEE-14 network) correspond to \emph{source vertices}, while we assume the dominant power flow pattern is active power flowing out of the very buses for simulation simplicity. All the other buses correspond to \emph{interior vertices}. Sink and source vertices are marked in red squares, and interior vertices are marked in black circles. So far we have applied our vertex classification method proposed in Section \ref{subsection-vertex-class} to the IEEE-14 test feeder.
\item Edge direction specification: Each transmission line between two buses corresponds to a directed edge. Edge directions are indicated by arrows, and edge weights are marked next to edges. Edge directions are determined by the `diodes’ positioning on the transmission lines, which are dictated by the attributes of the buses on two ends of the transmission line, i.e.,

\begin{enumerate}
\item In the case of connecting one sink vertex and one interior vertex, the diode faces toward the sink vertex. 
\item In the case of connecting one source vertex and one interior vertex, the diode faces toward the interior vertex.
\item In the case of connecting two interior vertices, the diode faces toward the interior vertex that is connected to the sink vertex.
\end{enumerate}

So far we have defined the incidence matrix $\mathcal{H}$ for the corresponding weighted directed graph using the proposed specification in Section \ref{subsection_weighted Laplacian matrix}, the numerical results of which are omitted due to page limit. 
\item Derivation of the weighted Laplacian: For the simplicity of the weighting diagonal matrix $B$, we assume all edge weights are $1$. So far we can derive the weighted Laplacian matrix $\mathcal{L}=\mathcal{H}_{o}B\mathcal{H}^T$ based on \emph{Definition} \ref{weightedLaplacian} for the corresponding graph.
\item Input profile: Since active power flows from buses with high voltage angles to buses with low voltage angles, we choose the angle profile conforming to the incidence matrix. We deliberately set each vertex angle to be a small shift from the reference angle $\alpha$. Phase shifts are in $[-0.6,0.6]$. The motivation of this treatment is that DC power flow essentially is a linearization of AC power flow, which takes the small angle difference as one of its several prerequisites. By limiting phase shifts in $[-0.6,0.6]$, we manage to keep angle differences smaller than 1.2. See the angle profile $\theta$ for the unreduced graph in the $3^{rd}$ column of Table~\ref{vertice_IEEE14}.
\item Output profile: we derive the vertex active power $P_{v}$ based on the expression given in (\ref{Pv=Ltheta}). See the vertex power $P_v$ in the $2^{nd}$ column of Table \ref{vertice_IEEE14}.
\item Reduction Stage \uppercase\expandafter{\romannumeral1}: we calculate the reduced weighted Laplacian matrix of the reduced graph preserving boundary vertices and vertices that are connected to boundary vertices via an edge using the expression given in (\ref{DefinitionKronreductiontoPF}). We then calculate the reduced power profile based on the expression given in (\ref{reduced-power-expression}). See the reduced power $P_v^{'}$ in the $4^{th}$ column of Table \ref{vertice_IEEE14}. The numerical results conform to \emph{Theorem} \ref{theorem-IO-of-lossless-PF}. See the reduced graph corresponding to the reduced weighted Laplacian matrix of Stage \uppercase\expandafter{\romannumeral1} in Fig. \ref{SINK1_SOURCE2_REDUC_1}. The successful delivery of the reduced directed graph conforms to \emph{Lemma} \ref{lemmageneralizedPF}.  
\item Reduction Stage 
\uppercase\expandafter{\romannumeral2}: we calculate the reduced weighted Laplacian matrix of the reduced graph eliminating all interior vertices based on the expression given in (\ref{DefinitionKronreductiontoPF}). We then calculate the reduced power profile based on the expression given in (\ref{reduced-power-expression}). See the reduced power $P_v^{''}$ in the $6^{th}$ column of Table \ref{vertice_IEEE14}, and the reduced graph corresponding to the reduced weighted Laplacian matrix of Stage \uppercase\expandafter{\romannumeral2} in Fig. \ref{SINK1_SOURCE2_REDUC_2}. The reduction results conform to \emph{Lemma} \ref{lemmageneralizedPF} and \emph{Theorem} \ref{theorem-IO-of-lossless-PF}.
\end{enumerate}

\begin{table*}
\renewcommand{\arraystretch}{1.3}
\caption{Vertex parameters of IEEE-14 test feeder before reduction (column 2,3), after stage \uppercase\expandafter{\romannumeral1} reduction (vertices $3,4,9,10,11$ eliminated) (column 4,5), and after stage \uppercase\expandafter{\romannumeral2} reduction (all interior vertices eliminated) (column 6,7)}
\centering
\begin{equation}
\begin{array}{ | c | c | c | c | c | c | c | }
\hline \text { Vertex }_{i} & P_{\text {vi }}(p.u.) & \theta_{i}\left({ }^{\circ}\right) &P_{\text {vi }}^{'}(p.u.) &\theta_{i}^{'}\left({ }^{\circ}\right)&P_{\text {vi }}^{''}(p.u.) & \theta_{i}^{''}\left({ }^{\circ}\right) \\ \hline
	1 & 0.58 & \alpha+0.5271 & 0.58 & \alpha+0.5271 & 1.98 & \alpha+0.5271 \\ \hline
	2 & 1.1 & \alpha+0.3371 & 1.6 & \alpha+0.3371 & \times & \times \\ \hline
	3 & 0.1 & \alpha-0.0629 & \times & \times & \times & \times \\ \hline
	4 & 0.1 & \alpha-0.1629 & \times & \times & \times & \times \\ \hline
	5 & 0.4 & \alpha+0.1371 & 0.6 & \alpha+0.1371 & \times & \times \\ \hline
	6 & 0.8 & \alpha+0.0371 & 1.1 & \alpha+0.0371 & \times & \times \\ \hline
	7 & 0.7 & \alpha+0.1371 & 1 & \alpha+0.1371 & \times & \times \\ \hline
	8 & 0.39 & \alpha+0.5271 & 0.39 & \alpha+0.5271 & 0.99 & \alpha+0.5271 \\ \hline
	9 & 0.1 & \alpha-0.2629 & \times & \times & \times & \times \\ \hline
	10 & 0.1 & \alpha-0.1629 & \times & \times & \times & \times \\ \hline
	11 & 0.1 & \alpha-0.0629 & \times & \times & \times & \times \\ \hline
	12 & 0.3 & \alpha-0.1629 & 0.3 & \alpha-0.1629 & \times & \times \\ \hline
	13 & 0 & \alpha-0.4629 & 0 & \alpha-0.4629 & 0 & \alpha-0.4629 \\ \hline
	14 & 0.1 & \alpha-0.3629 & 0.1 & \alpha-0.3629 & \times & \times \\ \hline

\end{array} \nonumber
\end{equation}
\label{vertice_IEEE14}
\end{table*}

\subsubsection{Results}

\begin{enumerate}
\item The lossless power flow network model of the IEEE-14 test feeder in a directed graph is successfully delivered.
\item The proposed weighted Laplacian matrix has been successfully derived from the directed graph, conforming to \emph{Definition} \ref{weightedLaplacian}.
\item The vertex angle input profile is suitably chosen to meet the linearization requirement.
\item The active power output profile is derived, showing that the proposed weighted Laplacian matrix functions as a mapping of system input to output.
\item Kron reduction is performed on the built lossless power flow network, of which the reduced results conform to \emph{Theorem} \ref{theorem-IO-of-lossless-PF}.
\item Notice that during stage \uppercase\expandafter{\romannumeral1}, $P_v^{'}$ of boundary vertices also remain unchanged after the transformation: $P_{v}^{'}=P_{v\alpha}+\mathcal{L}_{ac}P_{v{\alpha^c}}$. It will be interesting for future work to look into this matter.
\end{enumerate}

\subsection{Modified IEEE RTS-96 test system}
\subsubsection{Reduction process}
\begin{figure}[]
\centering
\includegraphics[width=1\textwidth]{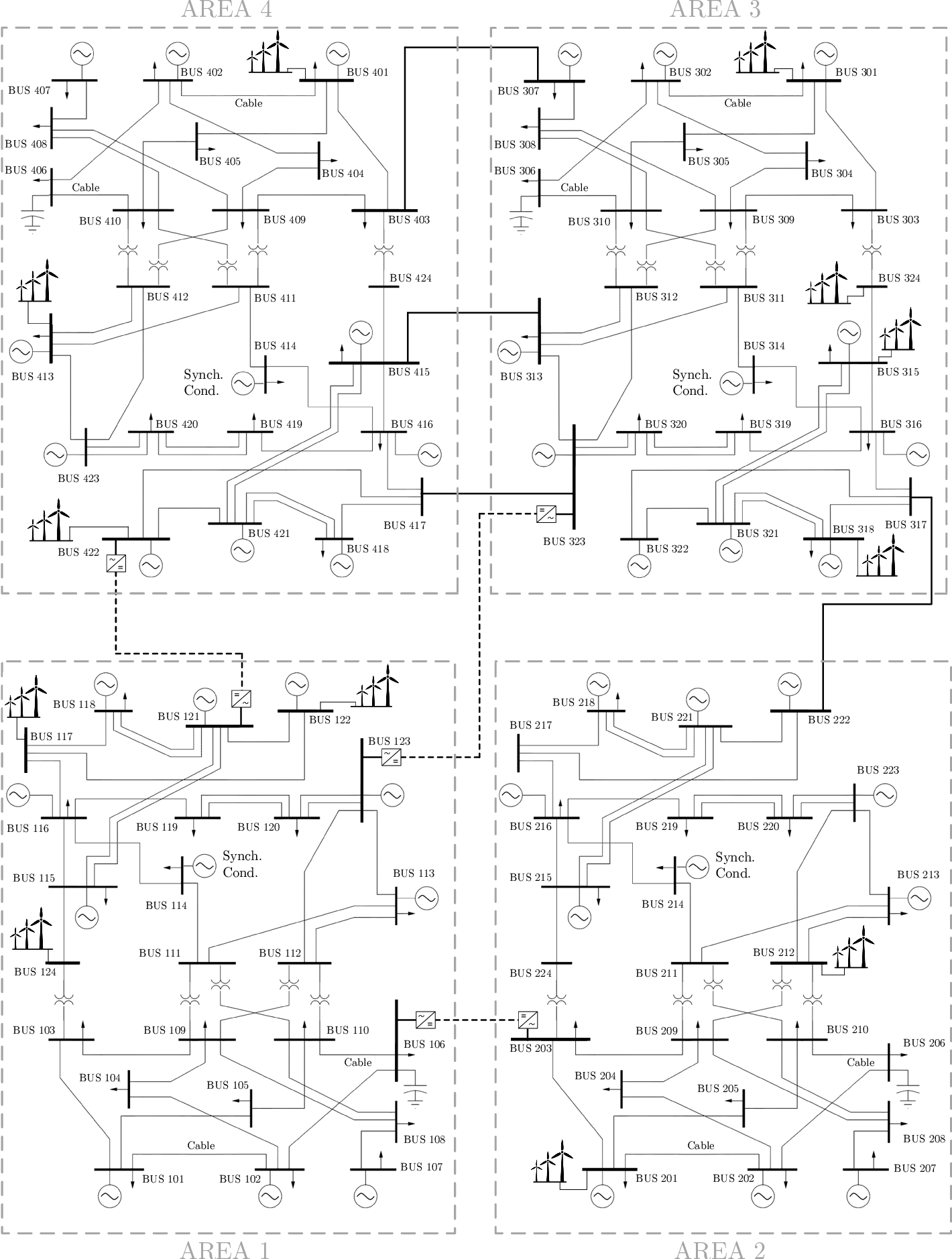}
\caption{Wiring diagram of the modified IEEE RTS-96 test system \cite{tosatto2019modified}}
\label{modifiedRTS96}
\end{figure}

\begin{figure}[]
\centering
\includegraphics[width=1\textwidth]{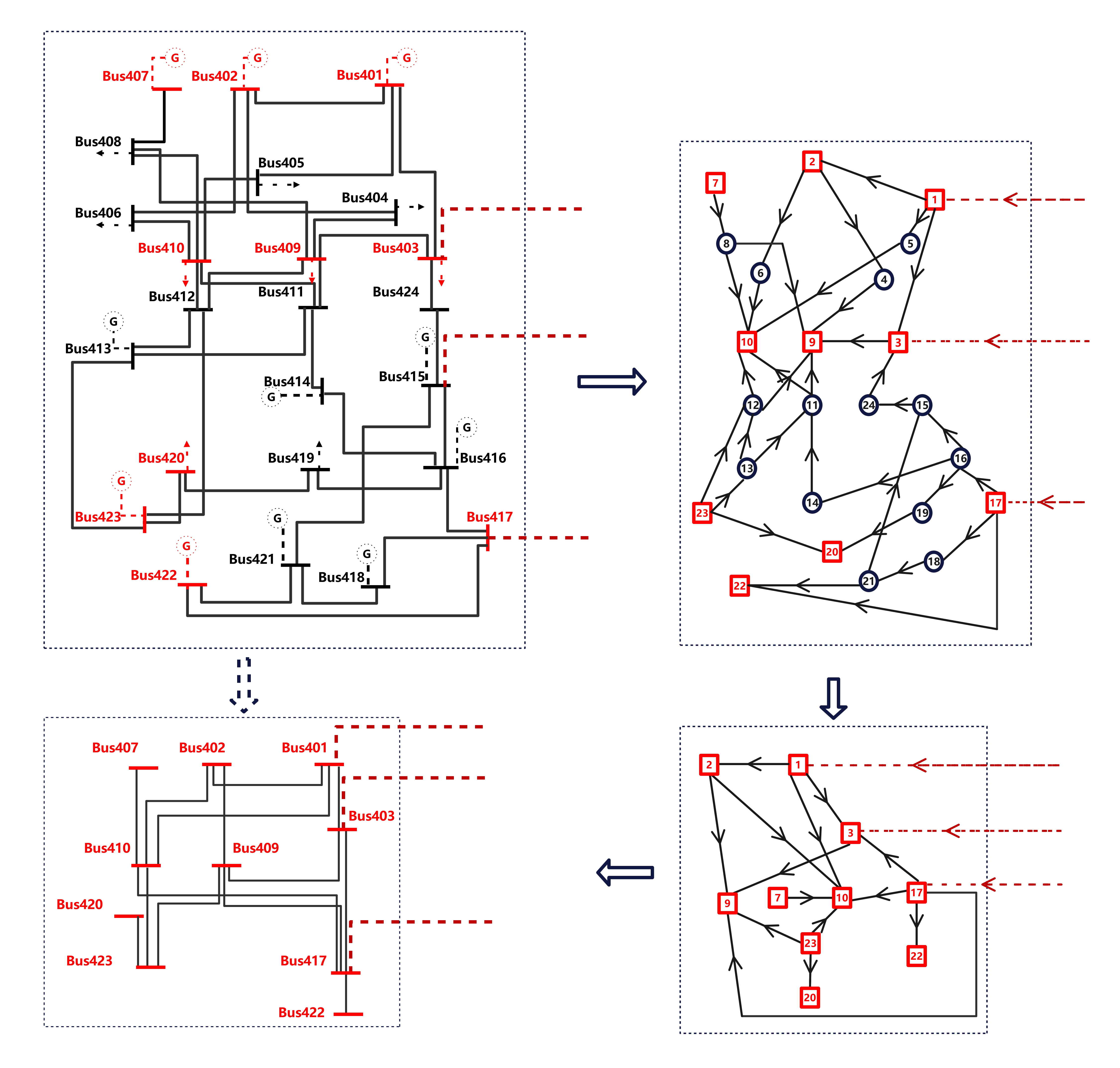}
\caption{Kron reduction on Area 4 of the modified IEEE RTS-96 test system}
\label{Kron_on_RTS96}
\end{figure}

In this example, we take Area $4$ of the modified IEEE RTS-96 test system from \cite{tosatto2019modified} in Fig.~\ref{modifiedRTS96} as the reduction object. Buses connected to generators, loadings, and buses in Area~3 are boundary vertices. The remaining vertices are interior vertices. The corresponding weighted Laplacian matrices of the original and the reduced graph are omitted due to the page limit. Bus angles and active power extractions are omitted as well. All interior vertices are eliminated during the reduction process in Fig.~\ref{Kron_on_RTS96}.
\subsubsection{Results}
\begin{enumerate}
\item The directed graph corresponding to Area $4$ of the IEEE RTS-96 test system is successfully derived.
\item The proposed weighted Laplacian matrix is derived based on the directed graph and is strictly equivalent to the conventionally defined Laplacian matrix, conforming to \emph{Definition} \ref{weightedLaplacian} and \emph{Theorem} \ref{1sttheorem}.
\item Kron reduced network is successfully derived by computing the Schur complement of the weighted Laplacian matrix.
\item The successful delivery of the Kron reduced network validates the scalability of the proposed reduction method.
\end{enumerate}~~

\section{\large Conclusions and recommendations}  \label{section-conclusionsandmore}
We have studied Kron reduction on directed graphs and on directed power flow networks. Our work was motivated by the gap in the existing research work between Kron reduction and its application to directed graphs, and the gap between Kron reduction and its application to electrical networks. We have proposed a novel formulation of the weighted Laplacian matrix for a directed graph in a way that the novel definition is strictly equivalent to the conventional definition of the weighted Laplacian.  We presented a comprehensive graph-theoretic analysis of Kron reduction to directed graphs. This analysis led to new physical insights regarding the application of power flow networks.
\par Our analysis demands answers to further questions, such as effective resistance and sensitivity analysis of Kron reduction to directed DC power flow networks, and Kron reduction to other characteristic electrical networks. Undirected/directed graphs with complex-valued weights as well for modelling power networks would be another interesting topic for future research.

\section*{\large Acknowledgment}
The first author would like to thanks EIT InnoEnergy and SENSE, for gaining the access to Europe’s largest innovation community, including top partners in business, research, and higher education. The paper presents research outcomes from the first author's MSc graduation project. The first author would like to thank the supervisor team (Dr. Zhiyong Sun and Prof. Siep Weiland) for stimulating conversations on this topic and for the guidance and effort in this project.  

%Bibliography
\bibliographystyle{unsrt}  
\bibliography{references}

\end{document}